\documentclass[epj]{svjour}
\usepackage{graphics}
\usepackage{color}      
\usepackage{braket}
\usepackage{pgf,tikz,pgfplots}
\usepackage{lscape}
\usepackage{rotating}
\usepackage{amsmath} 
\usepackage{amsfonts} 
\usepackage{amssymb}
\usepackage{amsfonts}
\usepackage{mathrsfs}
\usepackage[mathscr]{eucal}
\usepackage{etoolbox}
\usepackage{mathtools, cuted}
\usepackage{lipsum, color}
\usepackage{flushend}

\def \be  {\begin{equation}}
\def \ee  {\end{equation}}
\def \ee  {\end{equation}}
\def \bea {\begin{eqnarray}}
\def \eea {\end{eqnarray}}
\newcommand{\nn}{\nonumber}

\renewcommand{\figurename}{{\bf Fig.}}
\renewcommand{\tablename}{{\bf Tab.}}
\newcommand{\dslash}{\ensuremath{\partial\hspace{-1.2ex} /}}

\renewcommand{\figurename}{{\bf Fig.}}
\renewcommand{\tablename}{{\bf Tab.}}
\begin{document}
\title{Chiral magnetic properties of QCD phase--diagram}
\author{Abdel Nasser Tawfik\inst{1}\thanks{\emph{Present address:} Goethe University, Institute for Theoretical Physics (ITP), Max-von-Laue-Str. 1, D-60438 Frankfurt am Main, Germany}%
\and Abdel Magied Diab\inst{2}
}                     
%
%
\institute{Egyptian Center for Theoretical Physics (ECTP), Juhayna Square of 26th--July--Corridor, 12588 Giza, Egypt \and Modern University for Technology and Information (MTI),  Faculty of Engineering, 11571 Cairo, Egypt}
\date{Received: date / Revised version: date}
%
\abstract{
The QCD phase--diagram is studied, at finite magnetic field. Our calculations are based on the QCD effective model, the SU($3$) Polyakov linear--sigma model (PLSM), in which the chiral symmetry is integrated in the hadron phase and in the parton phase, the up--, down-- and strange--quark degrees of freedom are incorporated besides the inclusion of Polyakov loop potentials in the pure gauge limit, which are motivated by various underlying QCD symmetries. The Landau quantization and the magnetic catalysis are implemented.
The response of the QCD matter to an external magnetic field such as magnetization, magnetic susceptibility and permeability has been estimated. We conclude that the parton phase has higher values of magnetization, magnetic susceptibility, and permeability relative to the hadron phase.
Depending on the contributions to the Landau levels, we conclude that the chiral magnetic field enhances the chiral quark condensates and hence the chiral QCD phase--diagram, i.e. the hadron--parton phase--transition likely takes place, at lower critical temperatures and chemical potentials. 
\PACS{
      {11.10.Wx}{Chiral transition}   \and
      {25.75.Nq}{Quark deconfinement, quark--gluon plasma production, and phase transitions}  \and
      {98.62.En}{Electric and magnetic fields}
     } 
} 
\maketitle

\section{Introduction}
\label{intro}

Exploring the quantum chromodynamic (QCD) phase--diagram and studying the phase structures and the deconfinement phase--transitions of strongly interacting matter are among the fundamental issues in nuclear physics. Studying QCD matter in laboratory is one of the greatest challenges of the experimental nuclear physics as the parton matter is not directly accessible. There are different experimental methods implemented in accelerating and colliding ions (hadrons). Due to the center--of--mass energy of the collision achieved, different domains of the QCD phase--diagram could be drawn. The first prediction of end of the hadron domain, at high temperatures, was fomulated long time before the invention of QCD, where the partons are assumed as the effective degrees--of---freedom (dof), at temperatures larger than the Hagedorn temperature $T_H$ \cite{Hagedorn:1965st,Hagedorn:1984hz}. The hadron matter forms fireballs of new particles, which can again produce new fireballs. In 1975, Cabibbo proposed a QCD phase diagram in $T-n_B$ plane \cite{Cabibbo:1975ig}, where $T_H$ in the statistical Bootstrap model (SBM) \cite{Hagedorn:1965st} was interpreted as the critical temperature $T_c$ and is conjectured to be associated to second--order phase--transition into the deconfinement state. At large baryon density, $n_B$, a weak interaction between quarks and gluons - due to the asymptotic dof - has been recognized \cite{Collins:1974ky}. A Historical summary on the QCD phase diagram and its investigation in the heavy--ion collisions (HIC) experiments are available, for instance, in refs. \cite{Baym:2001in,fukushima2008chiral}.

In statistical physics, the phase transitions are defined as singularities or non--analyticity in the free energy as a function of thermodynamic quantities. In lattice QCD simulations, the corresponding partition functions are taken as functional integrals over compact groups described and evaluated in dependence on the temperature $T$, the chemical potential $\mu$, the volume $V$, the magnetic field $eB$, $\cdots$, etc.  

\begin{figure*}[htb]
\centering
\includegraphics[width=6.5cm,angle=-0]{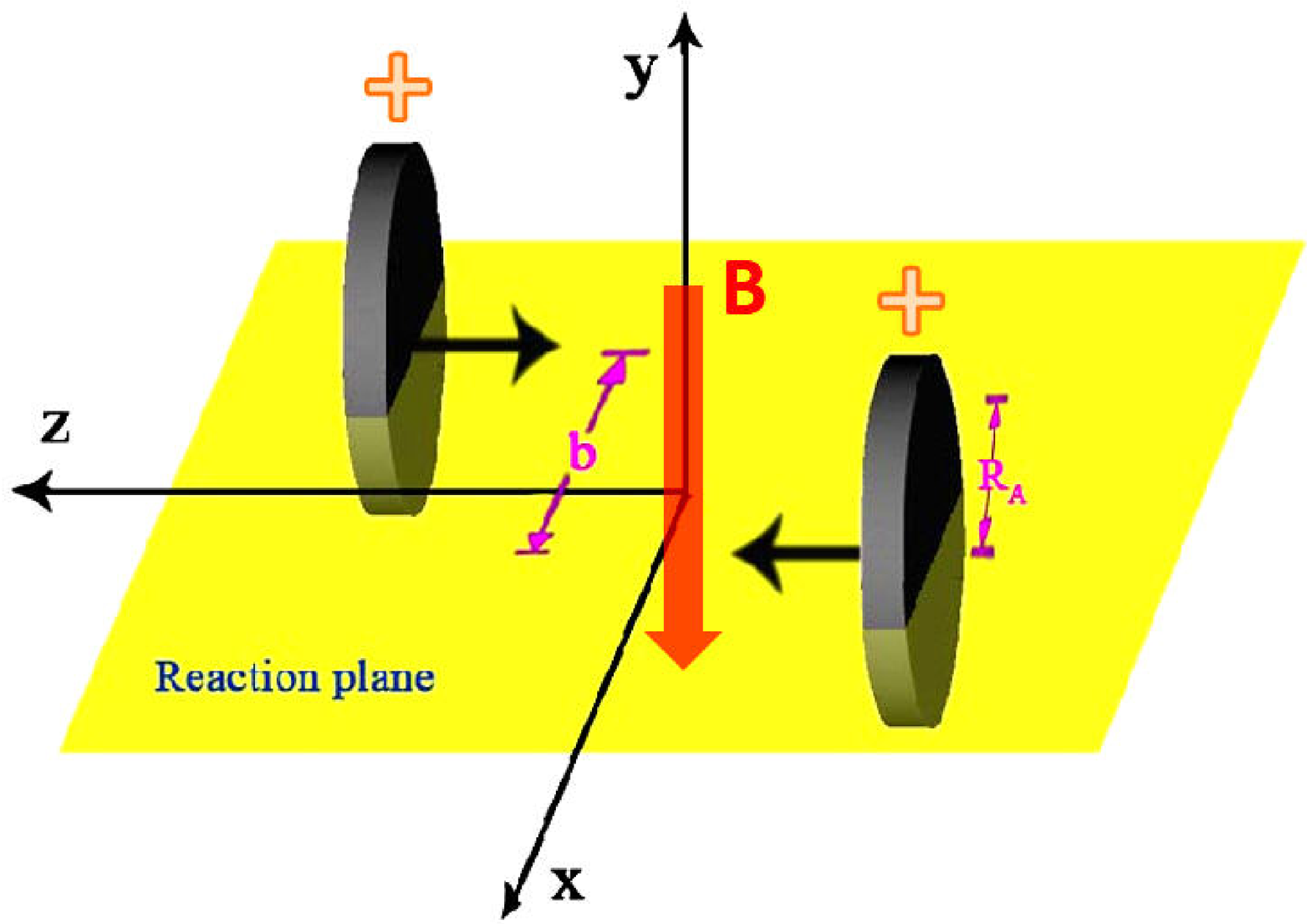}
\includegraphics[width=6.5cm,angle=-0]{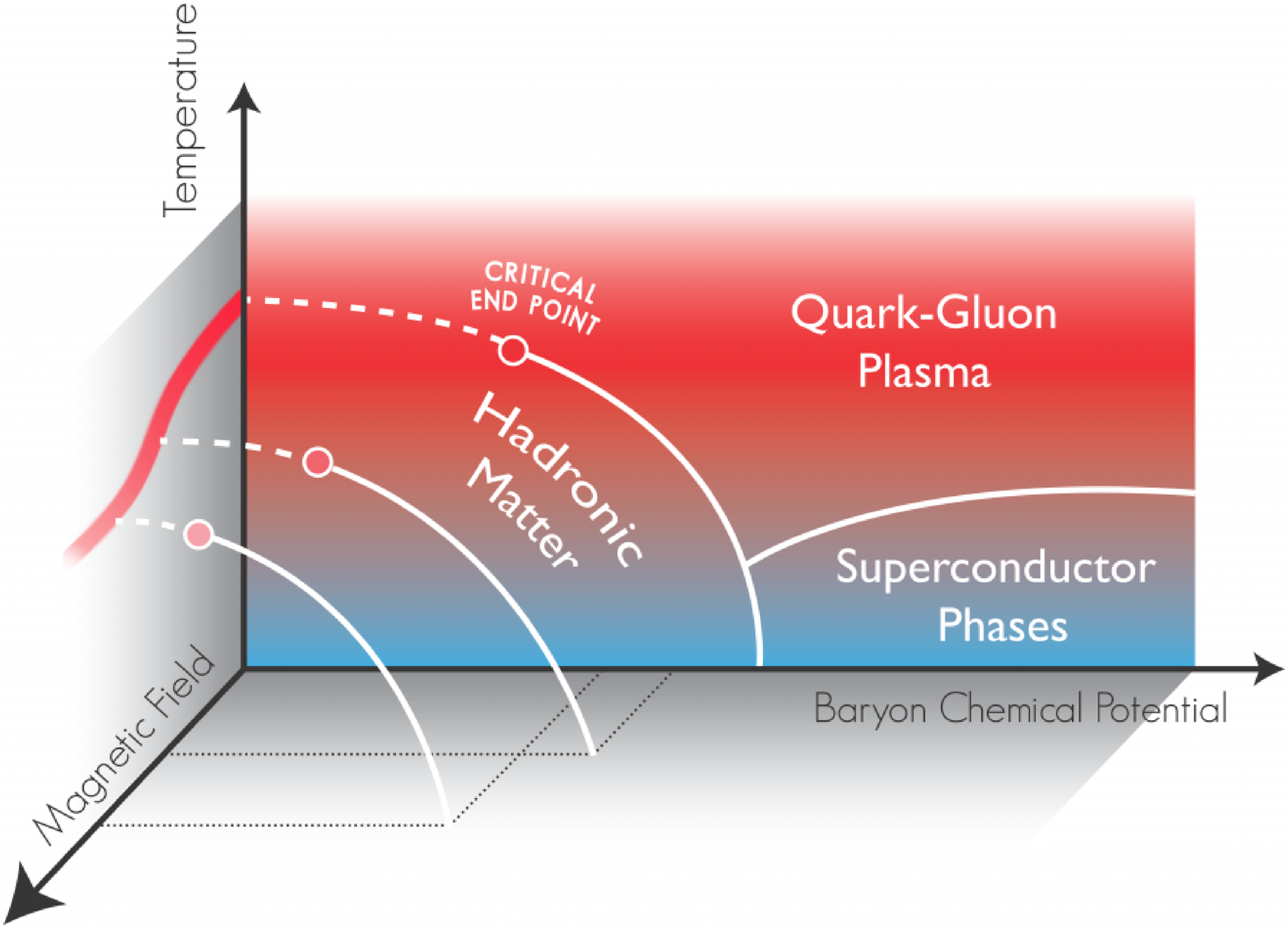}
\caption{Left panel shows the geometry of non--central HIC. $b$ is the impact parameter and $R_A$ is the radius of the nucleus. The magnetic field $B$ is expected to be perpendicular to the reaction plane due to the left--right symmetry of the collision geometry. The figure is taken from \cite{huang2016electromagnetic}. Right panel illustrates a schematic QCD phase--diagram. \label{fig:collisons0}
}
\end{figure*}

In HIC and due to oppositely ultra--relativistic motion of colliding heavy--ions, a huge magnetic field can be generated. Their motion generates an electric current which in turn induces magnetic field to the system. In a non--central HIC, the two counter--propagating nuclei collide, at finite impact parameter $b$. In Fig. \ref{fig:collisons0} the magnetic field in the center of the over--lapping surface in $Au+Au$ collisions, for instance, at $b=10~$fm and $\sqrt{s_{NN}}=200~$GeV, is visualized. This is perpendicular to the reaction plane owing to the symmetry of the collision.  Let us assume that all colliding nuclei are located, at the center of the nucleus, by applying Biot--Savart law,
\bea
- e B_y \sim 2 Z_{Au} \gamma \frac{e^2}{4\pi}\, v_z \left(\frac{2}{b}\right)^2 \approx 10^{19}\, \mbox{Gauss},
\eea
where the negative sign appears due to the assumption that the magnetic field is pointing in $-y$ direction, $v_z = (1-(2m_N/\sqrt{s})^2)^{1/2}\, \approx 0.99$ is the velocity of  accelerated nuclei, $m_N$ is the nucleon mass, $\gamma=1/\sqrt{1-v_z^2} \approx\, 100$ is the Lorentz gamma, and $Z_{Au}=79$ is the change number of the gold nucleus. It is assumed that such a huge magnetic field would have great impacts on the dynamics of the parton (quarks and gluons) matter produced in HIC. 

Over the last few decades, great efforts have been done to map out the QCD phase--diagram. Right panel of Fig. \ref{fig:collisons0} illustrates a schematic version. At low temperatures and baryon chemical potentials, the quarks and gluons are still confinement forming colorless hadrons. At $T\sim 150~$MeV, there is a crossover to the partonic colored phase; quark--gluon plasma. With increasing baryon--chemical potential, at low temperatures, the quarks shall be grouped in pairs known as correlated Cooper--pairs, which likely condense. Various experiments aim at studying the phase structures and the QCD phase--diagram such as RHIC at BNL $\sqrt{s_{NN}}=200~$GeV and LHC at CERN with energies up $\sqrt{s_{NN}}=13~$Tev. At $\sqrt{s_{NN}}=4-11~$GeV, the large density regime shall be explored by CBM at FAIR and MPD at JINR. 

In the heavy--ion experiments, as a result of the non--central heavy--ion collisions, a huge magnetic field could be created. For instance, at RHIC and LHC energies, the magnetic field ranges between $m_\pi^2$ and $10-15 m_\pi^2$, respectively \cite{Vachaspati:1991nm,bzdak2012event}, where $m_\pi^2 \sim 10^8~$Gauss.  It is worth mentioning that this value of course is just a snapshot, since the magnetic field is strongly time--dependent. Detailed discussion on such a dynamical system does not lay within the scope of the present study. On the other hand, the proposed approach, the PLSM, take into account the evolution of such dynamical system. The magnetic field lifetime in HIC including the electric and chiral magnetic effects \cite{McLerran:2013hla,Tuchin:2013apa} and the electromagnetic impacts on the heavy--ion phenomenology are reviewed in ref. \cite{Tuchin:2014hza}. The features of the electromagnetic fields in HIC shall be addressed, quantitatively. 

So far, there are various numerical approaches supporting the concept of magnetic catalysis in hot quark--matter and well agreeing with the recent PLSM result in strong magnetic--field such as the numerical lattice QCD simulations \cite{Bali:2011qj,Levkova:2013qda,Bonati:2013vba,Bruckmann:2013oba,Bali:2013txa}, the hadron resonance gas (HRG) model \cite{Endrodi:2013cs,Bhattacharyya:2015pra}, and the Polyakov--Nambu--Jona--Lasinio model \cite{Ratti:2005jh,Ferreira:2015gxa,Ferreira:2014kpa,Farias:2014eca,Chao:2013qpa,Mei:2020jzn,Mao:2016fha,Andersen:2021lnk}. Great details on understanding the phase structure of the QCD matter in strong magnetic--field are reviewed, for instance, in refs. \cite{Shovkovy:2012zn,Andersen:2014xxa,Kharzeev:2012ph,Gatto:2012sp}.

In 1960s, the linear sigma model (LSM), a low--energy model, was introduced by Gell--Mann and Levey \cite{gell1960axial}, long before the invertion of the Quantum Chromdynamics (QCD), the theory of the strong interaction. Many studies have been performed on LSM $\mathcal{O}(4)$  at  (non)--zero temperature \cite{Lenaghan:1999si,Petropoulos:1998gt} and for $N_f= 2,3,$ and $4$ quark flavors \cite{levy1967currents,hu1974chiral,geddes1980spin}. Moreover, the LSM is coupled with the Polyakov loop fields, known as the PLSM, to include the interaction and dynamics of the colored gluons. We have a solid term in developing the PLSM  to obtain reliable results. For instance, in estimating the features of the moments in thermal QCD medium \cite{Tawfik:2014uka}, obtaining the thermal spectrum for masses of meson states and QCD equation--s--tate (EoS) in thermal and dense QCD medium at (non)--zero magnetic backgrounds \cite{Tawfik:2014gga,Tawfik:2019rdd,Tawfik:2016lih}, Furthermore, the magnetic properties of the QCD matter such as the magnetization, magnetic susceptibility and  inverse magnetic catalysis could be estimated \cite{Tawfik:2016gye,AbdelAalDiab:2016rje,Tawfik:2017cdx}. We have also improved the PLSM to study the thermal structure of the transport properties, the bulk and the shear viscosity, the thermal and the electric conductivity of the QCD  matter \cite{Tawfik:2016edq,Tawfik:2016ihn}. The extension to $N_f=4$)--PLSM  \cite{Diab:2016iig,AbdelAalDiab:2018hrx} is an essential improvement in order to match with the recent lattice QCD simulations.

The present paper is divided into two main sections. Section \ref{sec1} summarizes the main features of the effective QCD approach, the Polyakov linear--Sigma model. Section \ref{sect2} outlines the results obtained.

\section{The Approach}
\label{sec1}

\subsection{Polyakov Linear--Sigma Model}

Assuming that the hadronic degrees--of--freedom (dof) are the colors of their quark constituents, various effective models relying on the chiral symmetry of QCD have been proposed \cite{Scavenius:2000qd}. The linear sigma model (LSM) implements the chiral symmetry in the hadronic sector and incorporates additional partonic (quarks) dof. We briefly summary the basic concepts of the SU($N_f$) Polyakov LSM (PLSM) in section  \ref{lagrangian}. Various approaches for the Polyakov loop potentials shall be discussed in section \ref{PLOYAKOV}. The inclusion of the magnetic effects in the mean field approximation by means of the Landau quantization shall be elaborated in section \ref{sec:llquant}.

\subsection{SU(N$_f$) Lagrangian \label{lagrangian}}

For LSM with SU($3$)$_L \times$ SU($3$)$_R$ and $N_f =2, \,3,\,4$ quark flavors and $N_c$ color dof, the symmetric LSM Lagrangian ${\cal L}_{chiral} = {\cal L}_f + {\cal L}_m$, where the fermionic part is given as  
\bea
\label{eq:quarkL}
\mathcal{L}_{f} = \bar{\psi}\left[i \dslash - g\; T_a\,\left(\sigma_a + i\,  \gamma_5\, \pi_a + \gamma_{\mu} V_a ^{\mu}+\gamma_{\mu}\gamma_{5} A_a^{\mu} \right)\,\right] \psi, \hspace*{5mm}
\eea
with $\mu$ is an additional Lorentz index \cite{Koch:1997ei}, $g$ is the Yukawa coupling of the quarks to the mesonic contributions ${\cal L}_{m}={\cal L}_{SP}+{\cal L}_{VA}+ {\cal L}_{Int}+\mathcal{L}_{U(1)_A}$ represented to ${\cal L}_{SP}$ scalars ($J^{PC}=0^{++}$) and pseudo--scalars ($J^{PC}=0^{-+}$), ${\cal L}_{VA}$ to vectors ($J^{PC}=1^{-}$) and axial--vectors ($J^{PC}=1^{++}$) mesons and $ {\cal L}_{Int}$ being the interaction between them. The Lagrangian of the anomaly term is given by $\mathcal{L}_{U(1)_A}$  
\cite{gell1960axial,gasiorowicz1969effective,Ko:1994en,Parganlija:2008jf,Pisarski:1994yp,Kovacs:2013xca,Tawfik:2014gga}.
\bea
\label{eq:Lagrangian}
\mathcal{L}_{SP}&=&\mathrm{Tr}(\partial_{\mu}\Phi^{\dag}\partial^{\mu}\Phi-m^2
\Phi^{\dag} \Phi)-\lambda_1 [\mathrm{Tr}(\Phi^{\dag} \Phi)]^2 \nn \\ 
&-&\lambda_2 \mathrm{Tr}(\Phi^{\dag}
\Phi)^2 + \mathrm{Tr}[H(\Phi+\Phi^{\dag})], \label{scalar nonets} \\ 
\mathcal{L}_{AV}&=&-\frac{1}{4}\mathop{\mathrm{Tr}}(L_{\mu\nu}^{2}+R_{\mu\nu}^{2}
)+\mathop{\mathrm{Tr}}\left[  \left( \frac{m_{1}^{2}}{2}+\Delta\right)  (L_{\mu}^{2}+R_{\mu}^{2})\right] \nn \\
&+&i \frac{g_{2}}{2} (\mathop{\mathrm{Tr}}\{L_{\mu\nu}[L^{\mu},L^{\nu}]\}+\mathop{\mathrm{Tr}}\{R_{\mu\nu}[R^{\mu},R^{\nu}]\}){\nonumber}\\
&+& g_{3}[\mathop{\mathrm{Tr}}(L_{\mu}L_{\nu}L^{\mu}L^{\nu}
)+\mathop{\mathrm{Tr}}(R_{\mu}R_{\nu}R^{\mu}R^{\nu})] \nn \\ &+&g_{4}
[\mathop{\mathrm{Tr}}\left(  L_{\mu}L^{\mu}L_{\nu}L^{\nu}\right)
+\mathop{\mathrm{Tr}}\left(  R_{\mu}R^{\mu}R_{\nu}R^{\nu}\right)
]{\nonumber}\\
&+&g_{5}\mathop{\mathrm{Tr}}\left(  L_{\mu}L^{\mu}\right)
\,\mathop{\mathrm{Tr}}\left(  R_{\nu}R^{\nu}\right)  \nn \\ &+& g_{6}
[\mathop{\mathrm{Tr}}(L_{\mu}L^{\mu})\,\mathop{\mathrm{Tr}}(L_{\nu}L^{\nu
})+\mathop{\mathrm{Tr}}(R_{\mu}R^{\mu})\,\mathop{\mathrm{Tr}}(R_{\nu}R^{\nu
})],\label{vector nonets}
\\
\mathcal{L}_{Int}&=&\frac{h_{1}}{2}\mathop{\mathrm{Tr}}(\Phi^{\dagger}\Phi
)\mathop{\mathrm{Tr}}(L_{\mu}^{2}+R_{\mu}^{2})+h_{2}%
\mathop{\mathrm{Tr}}[\vert L_{\mu}\Phi \vert ^{2} + \vert \Phi R_{\mu} \vert ^{2}]  \nn \\ &+&2h_{3}%
\mathop{\mathrm{Tr}}(L_{\mu}\Phi R^{\mu}\Phi^{\dagger}),\label{INT}
\\ 
\mathcal{L}_{U(1)_A} &=& c[\mathrm{Det}(\Phi)+\mathrm{Det}(\Phi^{\dag})]+c_0 [\mathrm{Det}(\Phi)-\mathrm{Det}(\Phi^{\dag})]^2  \nn \\ &+& c_1 [\mathrm{Det}(\Phi)+\mathrm{Det}(\Phi^{\dag})]\,\mathrm{Tr} [\Phi \Phi^{\dag}]. \hspace*{10mm}
 \label{eq:Lagrangian}
\eea
Equation (\ref{scalar nonets}) represents kinetic and potential terms for the scalar meson nonets. The third term - in this expression - gives the explicit symmetry breaking, which is  defined in Eq. (\ref{symmtery}). This part of the Lagrangian creates scalar and pseudo--scalar mesonic states defined in $\Phi$ nonets, Eq. (\ref{fieldmatrix}). Equation (\ref{vector nonets}) represents the vector meson nonets involving explicit symmetry breaking as given in the second term in Eq. (\ref{symmtery}). It is obvious that the $3 \times 3$ matrix of the vector meson nonets involves vector and axial--vector fields, Eq. (\ref{fieldmatrix}). Thus, this part creates vector and axial--vector mesonic states and expresses the interactions between (pseudo)--scalar and (axial)--vector as outlined in Eq. (\ref{INT}). Because of the explicit and spontaneous symmetry breaking, an anomaly term $\mathcal{L}_{U(1)_A}$ in SU(3)$_r \times$ SU(3)$_\ell$ is included in the effective Lagrangian. The parameters $c, c_0, c_1$ have to be determined, experimentally \cite{Parganlija:2012fy}. The first two terms in this Lagrangian approximate the original axial anomaly term \cite{Fariborz:2008bd,Rosenzweig:1979ay}, while the third term, which is proportional to the first term, is a mixed one. The first anomaly term, in which other terms are used to compare with other effects of different anomaly terms on the hadronic structure \cite{Kovacs:2013xca}, is the one taken into account in the present calculations. 

In order to repreoduce the related experimental results, the higher--order terms with local chiral symmetry have been included in \cite{Parganlija:2012fy}. It is worthy highlighting that $\mathcal{L}_{U(1)_A}$ symmetry is anomalous \cite{weinberg1975u} and known as the QCD vacuum anomaly \cite{weinberg1975u,Schaefer:2008hk}, i.e. broken by quantum effects. Without this term, a ninth pseudo--scalar Goldstone boson corresponding to the spontaneous breaking of the chiral U(3)$_{\ell} \times$ U(3)$_r$ symmetry likely unfold \cite{weinberg1975u,Schaefer:2008hk}. Therefore, the anomaly term is essential and the local chiral symmetry would not cause further numerical problems, at a mass scale of $1-2~$GeV \cite{Parganlija:2012fy}. The constraint terms are conjetured to have great influences \cite{Parganlija:2012fy}. Thus, it is assumed that the $\mathcal{L}_{U(1)_A}$ problem is effectively controlled by the inclusion of the $c$--term. Also, it should be noticed that $m^2$ (squared tree--level masses of mesons) and $m_{1}^{2}$ have contributions from the spontaneous symmetry breaking \cite{Parganlija:2012fy}.

The inclusion of scalar and vector meson nonets in the Lagrangian of PLSM is possible with a redefinition for the contra--covariant derivative of the quark--meson contributions, Eq. (\ref{derivative}), where dof of scalar $\Phi$ and vector $L^\mu $ and $R^\mu$ meson nonets are coupled to $A^{\mu}$, the electromagnetic field. Eqs. (\ref{vector}) and (\ref{axial}), the left-- and right--handed field strength tensors, respectively, represent self interaction between vector and axial--vector mesons $A^{\mu}$. Emerging from the globally invariant PLSM Lagrangian, the local chiral invariance requires that $g_1=g_2=g_3=g_4=g_5=g_6=g$ \cite{Parganlija:2012fy}
\bea
D^{\mu}\Phi & \equiv &\partial^{\mu}\Phi-ig_{1}(L^{\mu}\Phi-\Phi R^{\mu
})-ieA^{\mu}[T_{3},\Phi], \label{derivative} \\
L^{\mu\nu}  &  \equiv &\partial^{\mu}L^{\nu}-ieA^{\mu}[T_{3},L^{\nu}]-\left\{
\partial^{\nu}L^{\mu}-ieA^{\nu}[T_{3},L^{\mu}]\right\}, \label{vector} \\
R^{\mu\nu}  &  \equiv &\partial^{\mu}R^{\nu}-ieA^{\mu}[T_{3},R^{\nu}]-\left\{
\partial^{\nu}R^{\mu}-ieA^{\nu}[T_{3},R^{\mu}]\right\}.\label{axial} \hspace*{5mm}
\eea
It is obvious that $T_{a}=\hat{\lambda}_{a}/2$ with $a=0\dots 8$ are nine U($3$) generators, where $\hat{\lambda}_{a}$ are Gell--Mann matrices with fields $\Phi$ of $3\times 3$ complex matrix comprising of scalars $\sigma_{a}$ ($J^{PC}=0^{++}$), pseudo--scalars $\pi _{a}$ ($J^{PC}=0^{-+}$), $V_{a}^{\mu}$, vectors ($J^{PC}=1^{- -}$), and $A_{a}^{\mu}$ axial--vectors ($J^{PC}=1^{++}$) meson states, which are given as
\bea
\Phi = \sum_{a=0}^{8} T_{a}(\sigma _{a}+ i \pi _{a}),  \quad && \quad
L^{\mu} = \sum_{a=0} ^{8} \, T_{a}\, (V_{a}^{\mu}+A_{a}^{\mu}), \nn \\ 
R^{\mu}  = \sum_{a=0}^{8}\, T_{a}\, (V_{a}^{\mu}-A_{a}^{\mu}). \quad && \label{fieldmatrix}
\eea
The chiral symmetry is explicitly broken by 
\bea
H=\sum_{a=0}^{8} T_a h_a , \qquad &&
\Delta=\sum_{a=0}^{8} T_a \delta_a.
\label{symmtery}
\eea
 
A non--vanishing vacuum expectation value for $\Phi$,  $\langle \Phi \rangle = T_a \sigma_a $ breaks the chiral symmetry, spontaneously. Because of  the parity is not broken in the vacuum, there are no non--vanishing vacuum expectation values for fields $\pi _{a}$. In $U(3)_V \times U(3)_A$ symmetry, these patterns of the explicit symmetry breaking have been obtained as given in ref. \cite{Lenaghan:2000ey}.

Due to finite quark masses in the (pseudo)--scalar and (axial)--vector sectors, the breaking of U$(3)_{A}$ if $H_{0},\Delta_{0}\neq0$ and the symmetry breaking of $U(3)_{V}\rightarrow$ SU(2)$_{V}\times U(1)_{V}$ if $H_{8},\Delta_{8}\neq0$ \cite{Lenaghan:2000ey}. The symmetry breaking terms are originated from U($3$)$_{L}\times$ U($3$)$_{R}=$U($3$)$_{V} \times$ U($3$)$_{A}$. They are proportional to the matrices $H$ and $\Delta$, Eq. (\ref{symmtery}). The spontaneous chiral symmetry breaking is conjectured to take place in vacuum state. Therefore, a finite vacuum expectation value for $\Phi$ and $\bar{\Phi}$ are assumed to carry the quantum numbers of the vacuum, itself \cite{gasiorowicz1969effective}. Thus, the explicit symmetry breaking components (diagonal) $h_0$, $h_3$  and $h_8$ and $\delta_0$, $\delta_3$ and $\delta_8$ vanish \cite{gasiorowicz1969effective}, leading to extracting three finite condensates $\bar{\sigma_0}$, $\bar{\sigma_3}$ and $\bar{\sigma_8}$. On the other hand, $\bar{\sigma_3}$ breaks the isospin symmetry SU($2$) \cite{gasiorowicz1969effective}. To avoid this situation, we restrict ourselves to SU($3$). This can be $N_f=2+1$ \cite{Schaefer:2008hk} flavors. Correspondingly, two degenerate light (up-- and down--quarks) and one heavier strange--quark are assumed, i.e. $m_u=m_d \neq m_s$, where the violation of the isospin symmetry is neglected. This facilitates the choice of $h_a$ ($h_0 \neq 0$, $h_3=0$ and $h_8 \neq 0$) and for $\delta _a$  ($\delta_0 \neq 0$, $\delta_3 =0$  and $\delta_8 \neq 0$).

\subsection{Polyakov loops\label{PLOYAKOV}}

The LSM Lagrangian can be coupled to the Polyakov loops \cite{Schaefer:2008hk,Mao:2009aq},
\begin{eqnarray}
\mathcal{L}=\mathcal{L}_{chiral}-\mathbf{\mathcal{U}}(\phi, \phi^*, T), \label{plsm}
\end{eqnarray}
where the second term, $\mathbf{\mathcal{U}}(\phi, \phi^*, T)$, represents the effective Polyakov loop potential \cite{Polyakov:1978vu}. There are various proposals motivated by different underlying QCD symmetries in the pure gauge limit. These are different parameterizations reproducing first--order transition, at $T \sim 187~$MeV, $N_c=3$, and $N_f=2+1$ \cite{Ratti:2005jh,Roessner:2006xn,Fukushima:2008wg}.
\begin{itemize} \itemsep0em  
\item A simple choice based on Ginzburg--Landau ansatz \cite{Ratti:2005jh,Schaefer:2007pw}. The underlying $Z(3)$ center symmetry which is spontaneously broken should be conserved. Hence, an expansion in terms of the order--parameter can be expressed as
\bea
\label{eq:upoly}
 \frac{\mathcal{U}_{\mbox{poly}}}{T^{4}} &=& -\frac{b_2}{4} \left(\, |\phi|^2+|\phi^*|^2 \right) -\frac{b_3}{6}\, (\phi^3+\phi^{*3}) \nn \\ &+&\frac{b_4}{16} \left(|\phi|^2+|\phi^*|^2\right)^2
\eea
with the temperature--dependent coefficients $b_2(T) = a_0 + a_1 (T_0/T) + a_2 (T_0//T)^2 + a_3 (T_0/T)^3$. The parameters are estimated from the pure gauge lattice simulations, such that the equation of state and the Polyakov loop expectation values are well reproduced. The $\Phi$ terms are required to break the U($1$) symmetry of the remaining terms to Z($3$). $a_0=6.75$, $a_1=-1.95$, $a_2=2.625$, $a_3=-7.44$, $b_3=0.75$ and $b_4=7.5$ \cite{Ratti:2005jh} and deconfinement temperature $T_0=270~$MeV, can be used to estimate the pure gauge QCD thermodynamics and the Polyakov loop potential as functions of temperature.
%
\item An improved ansatz for the logarithmic form constrains $\phi$ and $\phi^*$ \cite{Roessner:2006xn}. 
\bea
\label{eq:ulog}
\frac{\mathcal{U}_{\mbox{log}}}{T^{4}}&=& -\frac{1}{2}\,a(T)  \phi^*  \phi  + b(T) \ln \Big[1-6  \phi^* \phi + 4\left(\phi^{3}+\phi^{*3}\right) \nn \\ &-& 3 \left(\phi^* \phi \right)^{2}\Big],
\eea 
with the temperature--dependent coefficients $a(T)=a_0 + a_1 (T_0/T) + a_2 (T_0/T)^2$ and $b(T) = b_3 (T_0/T)^3$. This potential is qualitatively consistent with the leading--order results from strong coupling expansion \cite{langelage2011centre}. Equation (\ref{eq:ulog}) diverges as $\phi^* \rightarrow 1$. This sets on limits to the Polyakov loop variables, i.e. they remain small, especially at high temperatures. $a_0,\, a_1,\, a_2$, and $b_3$ can be determined from lattice QCD simulations,  Tab. \ref{FitparameterPLOY}. Higher--order terms have been included  \cite{Lo:2013hla}, 
\bea
\frac{\mathbf{\mathcal{U}}_{\mathrm{PolyLog}}}{T^4} &=&  \frac{-a(T)}{2} \; \phi^* \phi + b(T)\; \ln \Big[1- 6\, \phi^* \phi  \nn \\ &+& 4 \,( \phi^{*3} + \phi^3) - 3 \,( \phi^* \phi)^2 \Big]   \nn \\ 
&+& \frac{c(T)}{2}\, (\phi^{*3} + \phi^3) + d(T)\, ( \phi^* \phi)^2. \label{LogPloy}
\eea 

It should be noticed that if $c(T)$ and $d(T)$ vanish, Eq. (\ref{LogPloy}) is reduced to Eq. (\ref{eq:ulog}). The various coefficients in Eq. (\ref{LogPloy}) are determined \cite{Lo:2013hla} $x(T) = (x_0 + x_1 \left(T0/T\right) + x_2 \left(T0/T\right)^2)(1+x_3 \left(T0/T\right) + x_4 \left(T0/T\right)^2)$ and $b(T)= b_0(T0/T)^{b_1} (1-e^{b_2 (T0/T)^{b_3}})$, 
where $x=(a,c,d)$. The different parameters are also summarized in Tab. \ref{FitparameterPLOY}. 

\item A potential inspired by a strong--coupling analysis \cite{Fukushima:2008wg} 
\bea
\label{eq:ufuku}
\frac{\mathcal{U}_{\mbox{Fuku}}}{T^4} &=& -\frac{b}{T^3}  \Big[\,54 \exp{(-a/T)} \phi^* \phi  \nn \\ &+& \ln \Big(1 - 6 \phi^* \phi  - 3 (\phi^* \phi )^2 + 4 (\phi^3 +\phi^{*3})\Big) \Big]. \hspace*{5mm}
\eea
The nearest--neighbor interaction is given in the first term, while the logarithm term is the Haar measure, Eq. (\ref{eq:ulog}). There are only two parameters, $a$ determines the deconfinement transition, i.e. the transition temperature in pure gauge theory and $b$ controls the mixing of the chiral and the deconfinement transition. At deconfinement temperature $T_0 \simeq 270$MeV, $a=664$MeV and $b = 196.2$MeV$^3$.
\end{itemize}

\begin{table*}[htb]
\caption{Fit parameters of logarithmic \cite{Roessner:2006xn} and polynomial--logarithmic Polyakov loop potentials \cite{Lo:2013hla} deduced from recent lattice QCD simulations. \label{FitparameterPLOY}}
 \centering
 \begin{tabular}{p{0.12\linewidth} p{0.12\linewidth}  p{0.12\linewidth} p{0.12\linewidth} p{0.12\linewidth} p{0.12\linewidth}} 
 \hline \hline
 Ref. \cite{Roessner:2006xn} & $a_0$ & $a_1$ & $a_2$ & $b_3$ \\ [0.5ex] 
 & $3.51$ & $-2.47$ &  $15.2$ & $-1.75$ \\
 \hline \hline
 Ref. \cite{Lo:2013hla} & $a_0$ & $a_1$ & $a_2$ & $a_3$ & $a_4$ \\
 & $-44.14$ & $151.4$ & $-90.0677$ & $2.77173$ & $3.56403$ \\
 \hline 
 & $b_0$ & $b_1$ & $b_2$ & $b_3$ & \\
 & $-0.32665$ & $5.85559$ & $-82.9823$ & $3.0$ & \\
 \hline 
 & $c_0$ & $c_1$ & $c_2$ & $c_3$ & $c_4$ \\
 & $-50.7961$ & $114.038$ & $-89.4596$ & $3.08718$ & $6.72812$ \\
 \hline 
 & $d_0$ & $d_1$ & $d_2$ & $d_3$ & $d_4$ \\
 & $27.0885$ & $-56.0859$ & $71.2225$ & $2.9715$ & $6.61433$ \\
 \hline  \hline
\end{tabular}
\end{table*} 
 In the present work, we have utilized the higher--order parameterization of the Polyakov loop fields based on the alternatively--improved extension of $\phi$ and $\phi^*$, for instance,  the polynomial--logarithmic Polyakov loop potentials Eq. (\ref{LogPloy}).

\subsection{Landau quantization}
\label{sec:llquant}

The quantity $2n+1-\sigma$ can be replaced by a summation over the Landau Levels $0\, \leq \nu \, \leq \nu_{max_f}$. The earlier is the Lowest Landau Level (LLL), while the latter stands for Maximum Landau Level (MLL), i.e. $\nu_{max}$. For the sake of completeness, we recall that $2-\delta_{0 \nu}$ represents degenerate Landau Levels and $\nu_{max_{f}}$ contributes to the maximum quantization number, i.e. $\nu_{max_{f}} \rightarrow \infty$). 
\begin{equation}
\nu_{max_{f}} = \left\lfloor  \frac{\tau _f ^2 - \Lambda ^2 _{QCD}}{2 |q_f| B } \right\rfloor, \label{MLL}
\end{equation} 
where the brackets represent floor of enclosed quantities and   the parameter $\tau$ is related to $\mu$, at varying $T$,  i. e. for results given in varying $T$, we should take into consideration $\tau \equiv \mu_f$. Concretely, when analyzing the results in thermal medium, $\mu_f$ is given by $\tau_f$, while when analyzing the results in dense medium, $T$ is given as function of $\tau$.

Equation (\ref{MLL}) refers to the contribution of single and double degenerates for the upper Landau levels. For the present study, we merely need to highlight that various works should have been analyzed \cite{Menezes:2009uc,Wen:2016atg,Avancini:2011zz,Boomsma:2009yk,Tawfik:2015tga}. Their results in MLL, $\nu_{max_f}$, at different temperatures and densities, for instance, can be proposed in terms of the medium parameters such as $\mu$, $T$ and $e B$. For the present calculations, we have to determine the maximum LL in order to infinity the contributions of MLL. We assume MLL as $\nu_{max_f}$, where $f \rightarrow \infty$. For details, interested readers are kindly advised to consult ref. \cite{Tawfik:2017cdx}.

At finite magnetic field, the dispersion relation contributes the Landau Level, so that 
\bea 
E_{B, f} (B) =\left(p_{z}^{2}+m_{f}^{2}+|q_{f}|(2n+1-\Sigma) B\right)^{1/2}.  \label{eq:moddisp}
\eea
Thus, the dispersion relation itself is modified by a quantization number, $n$, known as the Landau quantum number. $\sigma$ is related to the spin quantum number, $\Sigma=\pm S_z/2$ and $m_f (q_f)$ is the  quark flavor mass (charge). The present study counts for different contributions from Landau levels. The chiral condensates and deconfinement order--parameters shall be analyzed in a wide range of temperatures, baryon chemical potentials, magnetic fields, so that the chiral QCD phase--diagram could be mapped out in various directions.

A more challenging question is how the Landau quantization is fixed by the magnetic field, let us consider, for example, a Fermi sphere of quarks, where the system is considered as  discrete energy levels with respect to the momentum space. At magnetic field background the all spin directions are aligned though the transverse magnetic field, $\vec{B} = B\hat{e_z}$ {\it "polarization"}. With increasing the magnetic field, the order of Fermi energy increases and the energy levels are discretizied. This is called the {\it "Landau levels"}. Moreover, the phenomena of magnetic catalysis is mainly defined as an enhancement of the dynamical symmetry breaking by an external magnetic field. For instance, upon increasing the magnetic field tends to enhance or {\it ''catalysis''} the quark--antiquark condensate. This chiral condensate is strongly associated with the breaking of chiral symmetries and also creates masses. Apparently, this is magnetic catalysis \cite{Tawfik:2017cdx}.

The phenomena of (inverse)--magnetic catalysis are reviewed in great detail in ref. \cite{Shovkovy:2012zn,Kharzeev:2012ph}, where the  magnetic field enhances the spontaneous symmetry breaking and the chiral critical temperature decreases as increases the magnetic field strength, i.e. inverse magnetic catalysis.

\subsection{Mean--field approximation \label{subsec:mean field} }

For a spatially uniform system in a thermal equilibrium, at finite temperature $T$ and finite quark chemical potential $\mu_f$, where $f$ stands for $u, d$ and $s$ quarks, the partition function can be constructed. The grand canonical partition function governs the change in numbers of particles and antiparticles. Then, a path integral over quark, antiquark and meson fields leads to \cite{Schaefer:2008hk}
\bea
\mathcal{Z} &=& \mbox{Tr} \,\exp{\left[-(\hat{\mathcal{H}}-\sum_{f=u, d, s}
\mu_f \hat{\mathcal{N}}_f)/T \right]} \nn \\
&=&  \int\prod_a \mathcal{D} \sigma_a \mathcal{D} \pi_a \int
\mathcal{D}\psi \mathcal{D} \bar{\psi} \;  \exp \Big[ \int_x \Big(\hat{\mathcal{H}} -\mu_f \hat{\mathcal{N}} \nn \\ &+& \sum_{f=u, d, s} \mu_f \bar{\psi}_f \gamma^0 \psi_f \Big) \Big],
\eea
in which, we have abbreviated the space--time integration as
\bea
\int_x\equiv \int^{\beta}_0 d\tau \int_V d^3 x.
\eea
The integration runs over {\it imaginary time} $\tau=i\,t$ from $0$ to $\beta=1/T$. For a symmetric quark matter, uniform independent chemical potentials are imposed $\mu_f \equiv \mu_{u}=\mu_{d}=\mu_s$ \cite{Schaefer:2007pw,Schaefer:2008hk,Scavenius:2000qd}.  
\bea
\mu_u &=& \frac{\mu_B}{3} + \frac{\mu_I}{2} + \frac{1}{3} \mu_Y, \nn \\
\mu_d &=& \frac{\mu_B}{3} - \frac{\mu_I}{2} + \frac{1}{3} \mu_Y, \\ 
\mu_s &=& \frac{\mu_B}{3} - \frac{2}{3} \mu_Y, \nn
\eea
where $\mu_B$, $\mu_I$ and $\mu_Y$ are the baryon, isospin and hyper--charge chemical potentials, respectively. 

Converting the condensates $\sigma_0$ and $\sigma_8$ into a pure non--strange $\sigma_l$ and a pure strange $\sigma_s$ quark flavor leads to \cite{Kovacs:2006ym} 
\bea
\label{sigms}
\left( {\begin{array}{c}
\sigma_l \\
\sigma_s
\end{array}}
\right)=\frac{1}{\sqrt{3}} 
\left({\begin{array}{cc}
\sqrt{2} & 1 \\
1 & -\sqrt{2}
\end{array}}\right) 
\left({ \begin{array}{c}
\sigma_0 \\
\sigma_8
\end{array}}
\right).
\eea

Then, the purely mesonic potential is given as
\begin{eqnarray}
U(\sigma_l, \sigma_s) &=& - h_l \sigma_l - h_s \sigma_s + \frac{m^2}{2} (\sigma^2_l+\sigma^2_s) - \frac{c}{2\sqrt{2}} \sigma^2_l \sigma_s  \nn \\ &&
+ \frac{\lambda_1}{2}  \sigma^2_l \sigma^2_s +\frac{(2 \lambda_1 +\lambda_2)}{8} \sigma^4_l  + \frac{(\lambda_1+\lambda_2)}{4}\sigma^4_s. \hspace*{6mm}
\label{pure:meson}
\end{eqnarray}
The quarks and anti--quark contributions can be divided into two regimes: 
\begin{itemize}
\item At vanishing magnetic field ($eB=0$) but finite temperature ($T$) and baryon chemical potential ($\mu_f$), 
\begin{eqnarray} 
&&\Omega_{\bar{q}q}(T, \mu _f) = -2 T \sum_{f=l, s} \int_0^{\infty} \frac{d^3\vec{p}}{(2 \pi)^3} 
\times \nn \\
&& \left\{ \ln \left[ 1+3\left(\phi+\phi^* e^{-\frac{E_f-\mu _f}{T}}\right) e^{-\frac{E_f-\mu _f}{T}}+e^{-3 \frac{E_f-\mu _f}{T}}\right] \right.  
\nonumber \\ &&  \left. +\ln \left[ 1+3\left(\phi^*+\phi e^{-\frac{E_f+\mu _f}{T}}\right) e^{-\frac{E_f+\mu _f}{T}}+e^{-3 \frac{E_f+\mu _f}{T}}\right] \right\}, \nn \\
&& \label{PloykovPLSM}
\end{eqnarray}

where the $E_f = [p^2 + m_f^2]^{1/2}$ is the flavor--dependent single--particle energies of quark mass $(f=l,s,c)$ as,
\bea
m_{l=u,d} = g \frac{\sigma_l}{2} \quad  m_{s} = g \frac{\sigma_s}{\sqrt{2}}   \quad m_{s} = g \frac{\sigma_c}{\sqrt{2}},
\eea
where the $\sigma$--field coupled through Yukawa coupling $g$ and the subscript $l$ being the degenerate light up and down quarks. The SU($3$) Polyakov linear--sigma model (PLSM) in mean--field approximation is utilized in analyzing the thermodynamic properties of quark matter in thermal and dense QCD medium at finite isospin asymmetry in  \cite{Tawfik:2019tkp}.
\item At finite  magnetic background ($eB\neq 0$) where the magnetic field $\vec{B} = B \hat{e_z}$, all the spin directions should be aligned in parallel with the magnetic field, and  the momentum directions are only determined according to chirality. The concepts of Landau quantization and magnetic catalysis, where the magnetic field is assumed to be oriented along the $z$--direction \cite{Fukushima:2008xe,Fukushima:2009ft}. One can notice that the spin (Polarization) plays an essential role in the dimensional reduction of Dirac particles such as quarks. This latter process is known as dimension reduction or magnetic catalysis effect \cite{Shovkovy:2012zn,Kharzeev:2012ph}. Moreover, the thermodynamic potential should be implemented in a finite temperature, $T$, chemical potential, $\mu_B$, and $B$ as,

\begin{eqnarray} 
&&\Omega_{\bar{q}q}(T, \mu _f, B) = - \sum_{f=l, s} \frac{|q_f| B \, T}{(2 \pi)^2}   \sum_{\nu = 0}^{\nu _{max_{f}}} (2-\delta _{0 \nu }) \int_0^{\infty} dp_z  
\times \nn\\
&&\left\{ \ln \left[ 1+3\left(\phi+\phi^* e^{-\frac{E_{B, f} -\mu _f}{T}}\right) e^{-\frac{E_{B, f} -\mu _f}{T}} +e^{-3 \frac{E_{B, f} -\mu _f}{T}}\right] \right. \nonumber \\ 
&& \left.+\ln \left[ 1+3\left(\phi^*+\phi e^{-\frac{E_{B, f} +\mu _f}{T}}\right) e^{-\frac{E_{B, f} +\mu _f}{T}}+e^{-3 \frac{E_{B, f} +\mu _f}{T}}\right] \right\}. \nn \\ &&  \label{PloykovPLSMeB}
\end{eqnarray}
 Accordingly, one can solve the Dirac equation to obtain the modified dispersion relation, Eq. {\ref{eq:moddisp}}.  At finite volume, $V$, and finite magnetic field, $eB$, the free energy is defined as  \hbox{$\mathcal{F} =-T \cdot \log [\mathcal{Z}]/V$} or 
\begin{eqnarray}
\mathcal{F} &=&  U(\sigma_l, \sigma_s) +\mathbf{\mathcal{U}}(\phi, \phi^*, T) + \Omega_{\bar{q}q}(T, \mu_f, B) \nn \\ &+& \delta_{0,eB} \,\Omega_{ \bar{q}q}(T, \mu _f),  \label{potential}
\end{eqnarray}
in which,the  LSM mesonic potential $U(\sigma_l, \sigma_s)$ counts for the contributions of the valence quarks. At very low temperatures, this part of the potential could be excluded, especially in the regime of temperatures typical for the QCD phase transition \cite{Tawfik:2014uka}
\begin{eqnarray}
\mathcal{F} &=&  \mathbf{\mathcal{U}}(\phi, \phi^*, T) + \Omega_{\bar{q}q}(T, \mu_f, B)  + \delta_{0,eB} \Omega_{\bar{q}q}(T, \mu_f).  \hspace*{6mm} \label{potential}
\end{eqnarray}
At vanishing magnetic field $B=0$, the second term vanishes, Eq. (\ref{PloykovPLSMeB}), while $\delta=1$ in the third term, i.e. $\Omega_{\bar{q}q}(T, \mu _f)$, Eq. (\ref{PloykovPLSM}), should be evaluated with the standard dispersion--relation, $E_f$. Therefore, Eq. (\ref{potential}), can be given as 
\begin{eqnarray}
 \mathcal{F}  
&=&  \mathbf{\mathcal{U}}(\phi, \phi^*, T) + \Omega_{\bar{q}q}(T, \mu _f). \label{potential1}
\end{eqnarray}
When the magnetic field is switched on, $\delta=0$ and therefore the fourth term vanishes, as well. In this case, Eq. (\ref{potential}) can be reduced to 
\begin{eqnarray}
\mathcal{F} &=&  \mathbf{\mathcal{U}}(\phi, \phi^*, T) + \Omega_{\bar{q}q}(T, \mu_f, B). \label{potential2}
\end{eqnarray}
\end{itemize}

The potential of quark and antiquark contribution, at finite magnetic filed, $\Omega_{\bar{q}q}(T, \mu_f, B)$, can be divided into two regimes as shown in Eq. (\ref{PloykovPLSMeB}):
\begin{enumerate}
\item The first integral in Eq. (\ref{PloykovPLSMeB}) refers to the contribution of the magnetic field in zero Landau level,  $\nu=0$. In this case, the modified dispersion--relation, Eq (\ref{eq:moddisp}), tends to the standard one, $E_f$.
\item The second integral in Eq. (\ref{PloykovPLSMeB}) gives the contribution of the magnetic field for the upper Landau levels, $\nu=1\rightarrow \infty$, where the dispersion relation will be modified, $E_{B,f}$, as given in Eq. (\ref{eq:moddisp}).
\end{enumerate}
Furthermore, the excluding of $U(\sigma_l, \sigma_s)$ potential from the free energy,  especially, at high temperature, tends to maintain dominant contributions of the sea quarks than the valence quarks. In order to evaluate the expectation values of the PLSM order--parameters
\begin{eqnarray}\label{cond1}
\left. \frac{\partial \mathcal{F}}{\partial \sigma_l}= \frac{\partial
\mathcal{F}}{\partial \sigma_s}= \frac{\partial \mathcal{F}}{\partial
\phi}= \frac{\partial \mathcal{F}}{\partial \phi^*}\right|_{min} =0.
\end{eqnarray}
In nonzero chemical--potential ($\mu\ne0$) and finite Ployakov loop variables, the PLSM free energy, at finite $V$,  Eq. (\ref{potential}), is complex. A minimization of a such function would be seen as void of meaning.

An analysis of the order parameters is given by minimizing the real part of the PLSM free energy ($\mbox{Re}\;  \mathcal{F}$). In principle, the (thermal) expectation values of the Ployakov loop $\bar{\phi}$ and its conjugate $\bar{\phi}^*$ must be real quantities as discussed in ref. \cite{Dumitru:2005ng}. The solutions of these equations can be determined by minimizing the real part pf the PLSM free energy ($\mbox{Re}\;  \mathcal{F}$) at a saddle point. The remaining parameters are the chiral order--parameters $\bar{\sigma_l}$, $\bar{\sigma_s}$ and the Polyakov--loop expectation values $\bar{\phi},\; \bar{\phi}^*$ as functions of $T$, $\mu$ and $eB$.

\section{Results}
\label{sect2}

\subsection{Chiral and deconfinement order--parameters \label{Order:parameters}}

The estimation of the chiral condensates ($\sigma_l$ and $\sigma_s$) and the deconfinement order--parameters ($\phi$ and $\phi^*$) in dense and thermal medium should be first computed by minimizing the free energy, Eq. (\ref{cond1}). The parameters of PLSM, as discussed in sections \ref{lagrangian} and \ref{PLOYAKOV}, are estimated, at mass of (vacuum) sigma--meson $\sigma=800~$MeV, are measured (vacuum) light and strange chiral condensates are $\sigma_{l_o}=92.5~$ MeV and $\sigma_{s_o}=94.2~$MeV, respectively.

\begin{figure*}[htb] 
\centering
\includegraphics[width=4cm,angle=-90]{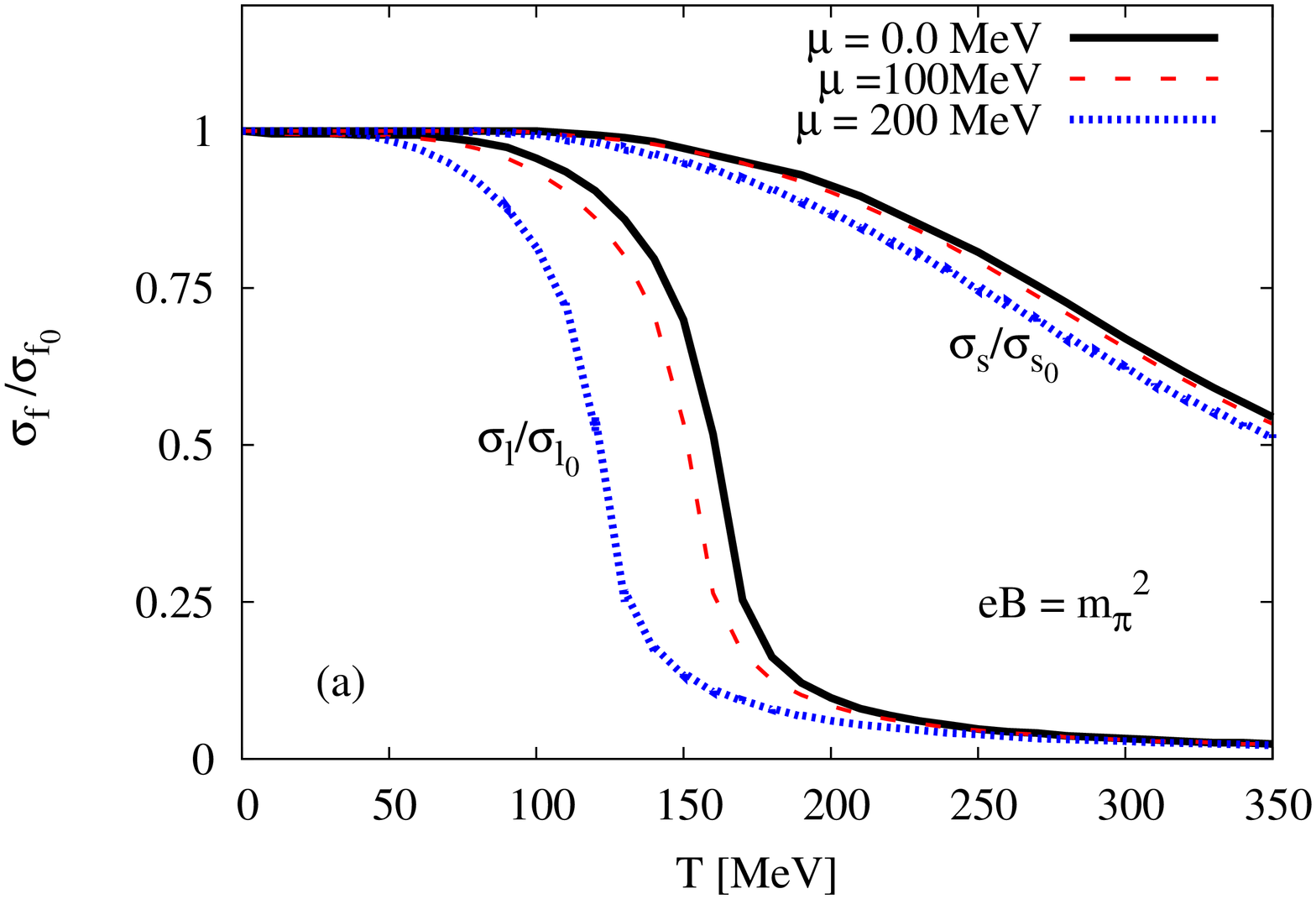}
\includegraphics[width=4cm,angle=-90]{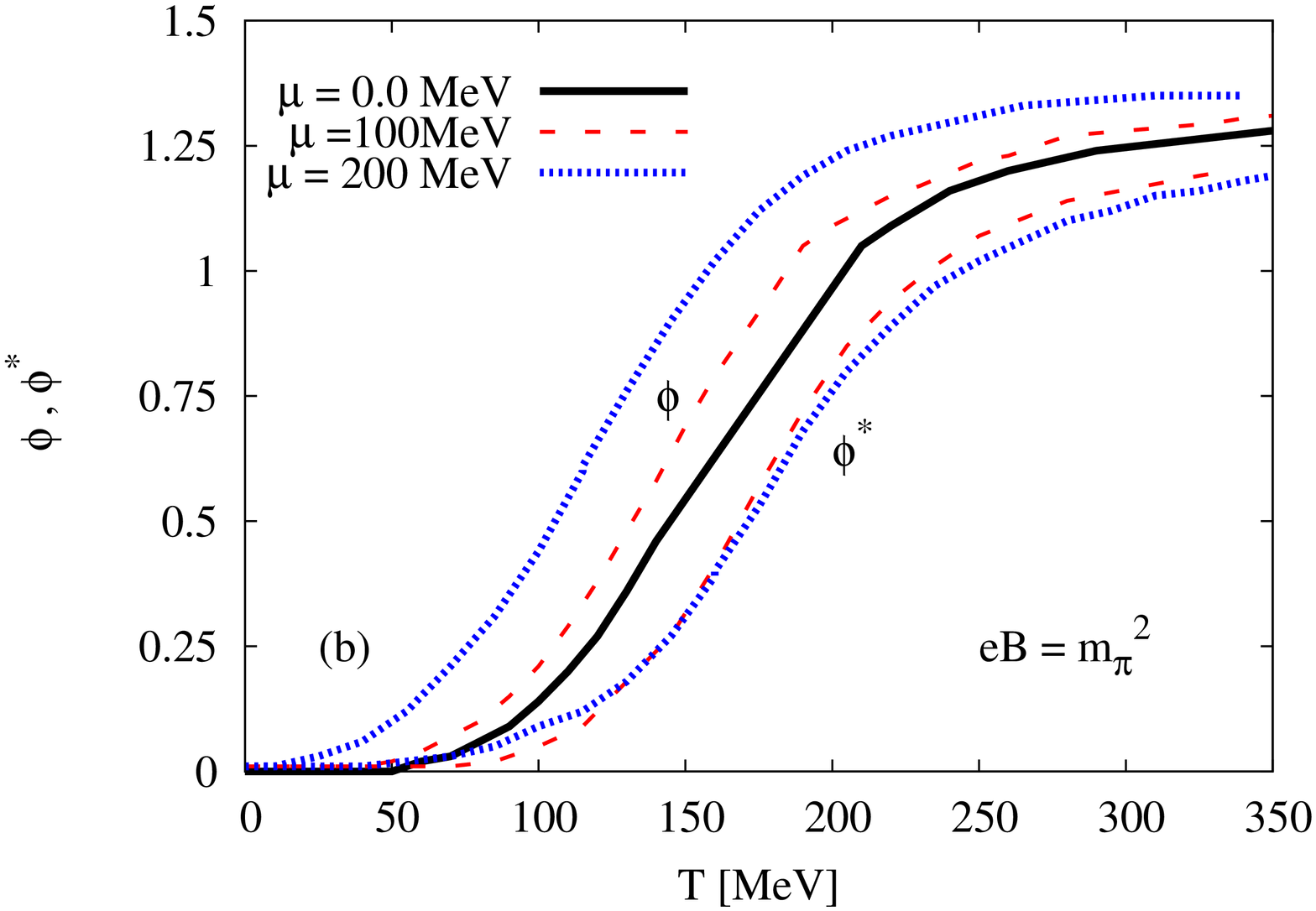}\\
\includegraphics[width=4cm,angle=-90]{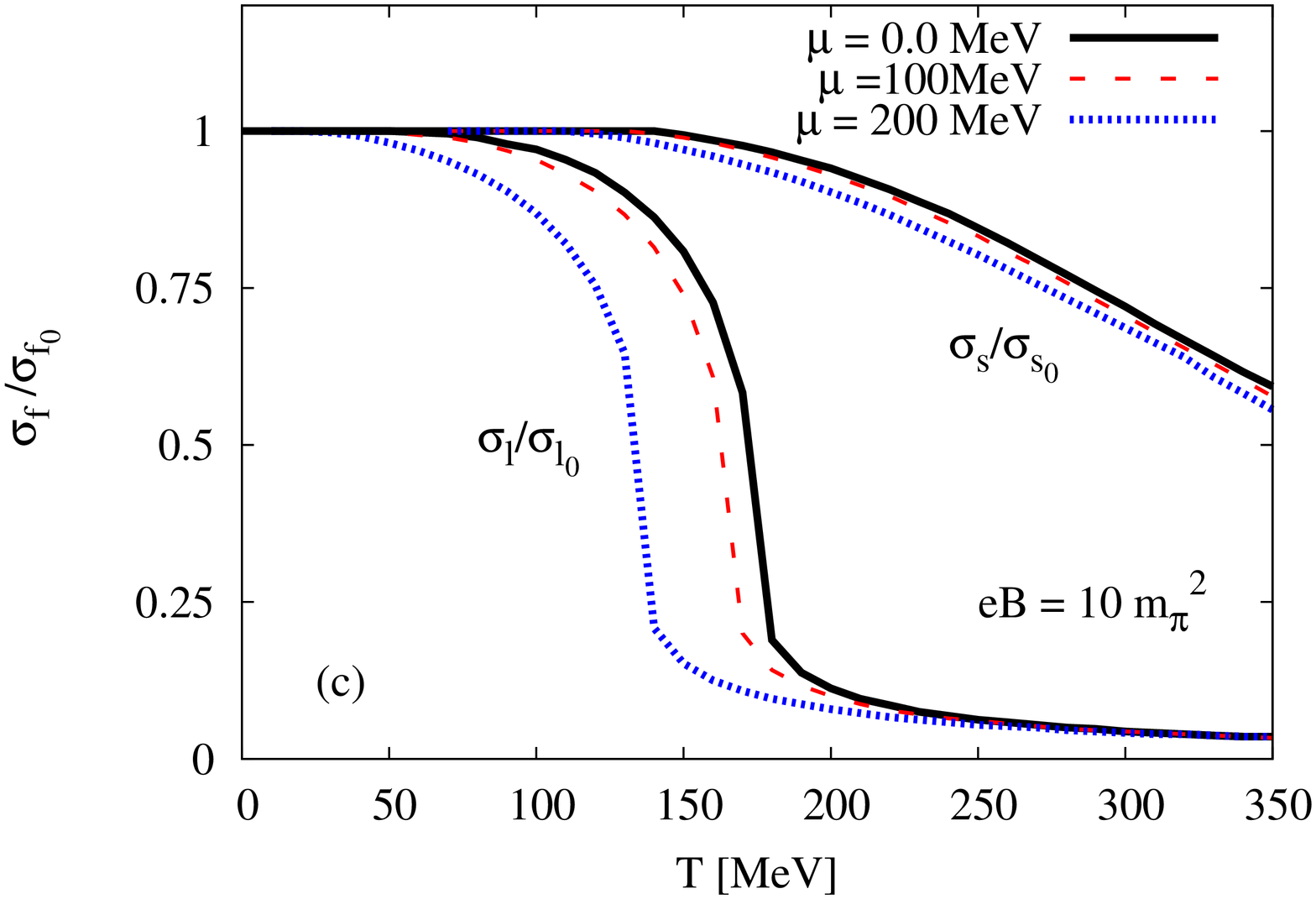}
\includegraphics[width=4cm,angle=-90]{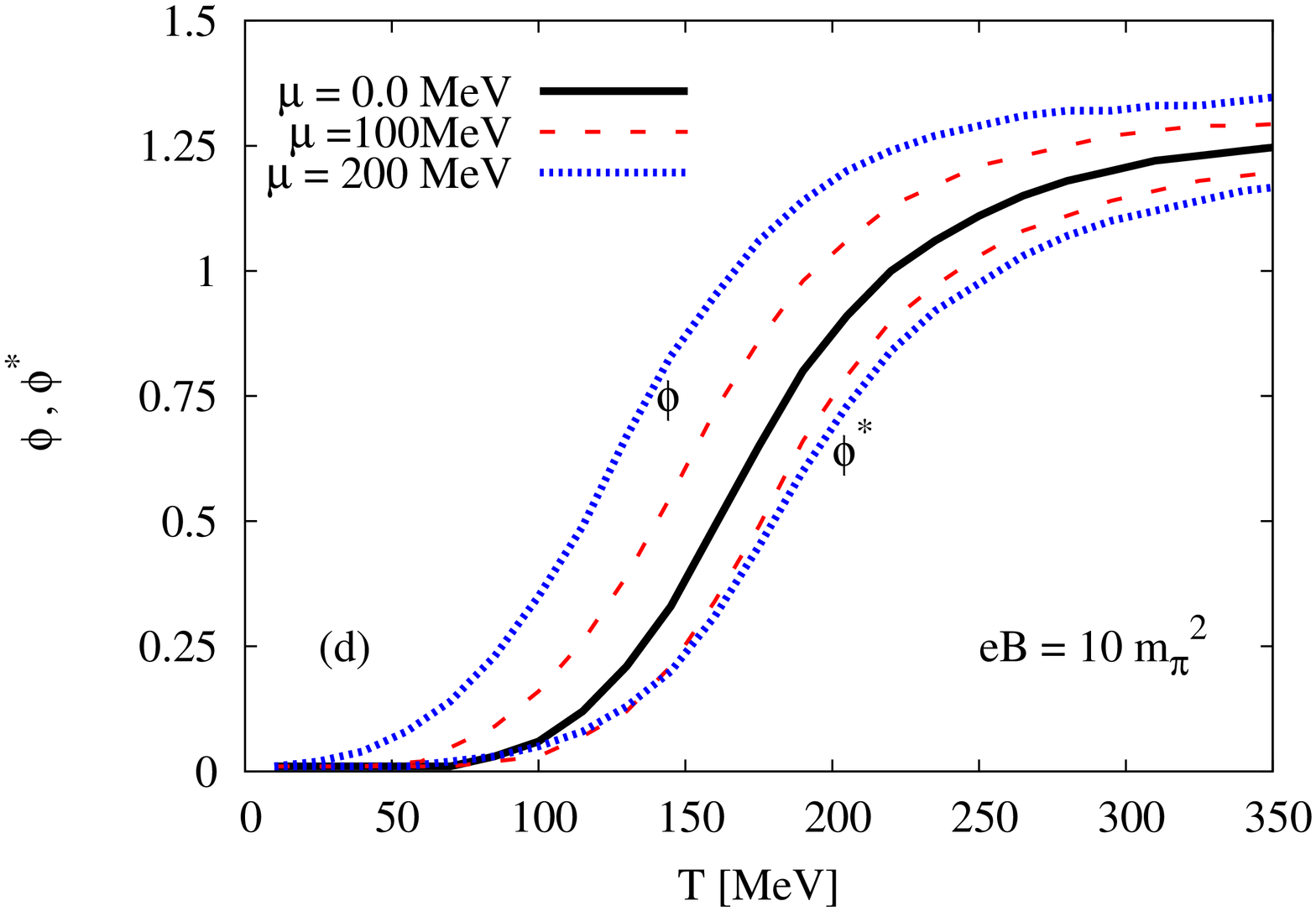}
\caption{Left--hand panels (a) and (c) show normalized chiral condensate with respect to the vacuum value as functions of temperature. Right--hand panels (b) and (d) give the expectation values of the Polyakov loop fields ($\phi$  and $\phi^*$) as functions of temperatures, as well. The upper panels presents the PLSM results, at magnetic fields $eB =m^2_{\pi}~$, while the bottom panels at $eB=10 \,m^2_{\pi}~$ and $\mu=0$ (solid curves), $100$ (dashed curves), $200~$MeV (dotted curves).  \label{fig:sbtrc1}}
\end{figure*}

Figure \ref{fig:sbtrc1} presents both PLSM chiral--condensates and order--parameters, at finite baryon chemical potentials and magnetic fields $eB=m^2_{\pi}$ (upper panel) and $eB=10\, m^2_{\pi}~$ (bottom panel) as functions of $T$. In the left--hand panels, (a) and (c) for normalized chiral--condensates, $\sigma_{l}/\sigma_{l_o}$ and $\sigma_{s}/\sigma _{s_o}$, we notice that the chiral condensates are shifted to lower values with increasing baryon chemical potential. This means that the chiral critical temperature ($T_{\chi}$) decreases with increasing $eB$ and with increasing $\mu$, as well. Therefore, we can draw a conclusion that the effect of strong magnetic field is almost the same as that of the baryon chemical potentials, namely on decreasing the phase transition (crossover). 

In right--hand panels (b) and (d) for the Polyakov loop parameters, $\phi$ and $\phi^*$ as functions of $T$, at diffident baryon chemical potentials, we find that both order parameters become differentiable, at finite $\mu$; $\mu=100~$MeV (dashed curves) and $\mu= 200~$MeV (dotted curves), while at $\mu=0~$MeV (solid curves), $\phi=\phi^*$. The deconfinement critical temperature, $T_{\phi}$, is shifted to lower values as the magnetic field and baryon chemical potential increase. We thus draw the conclusion that the $\phi$ and $\phi^*$ have opposite dependence on temperatures. The latter increases as the magnetic field and the baryon chemical potential $\mu$ increase, while the earlier decreases.

\begin{figure*}[htb]
\centering{
\includegraphics[width=5cm,angle=-90]{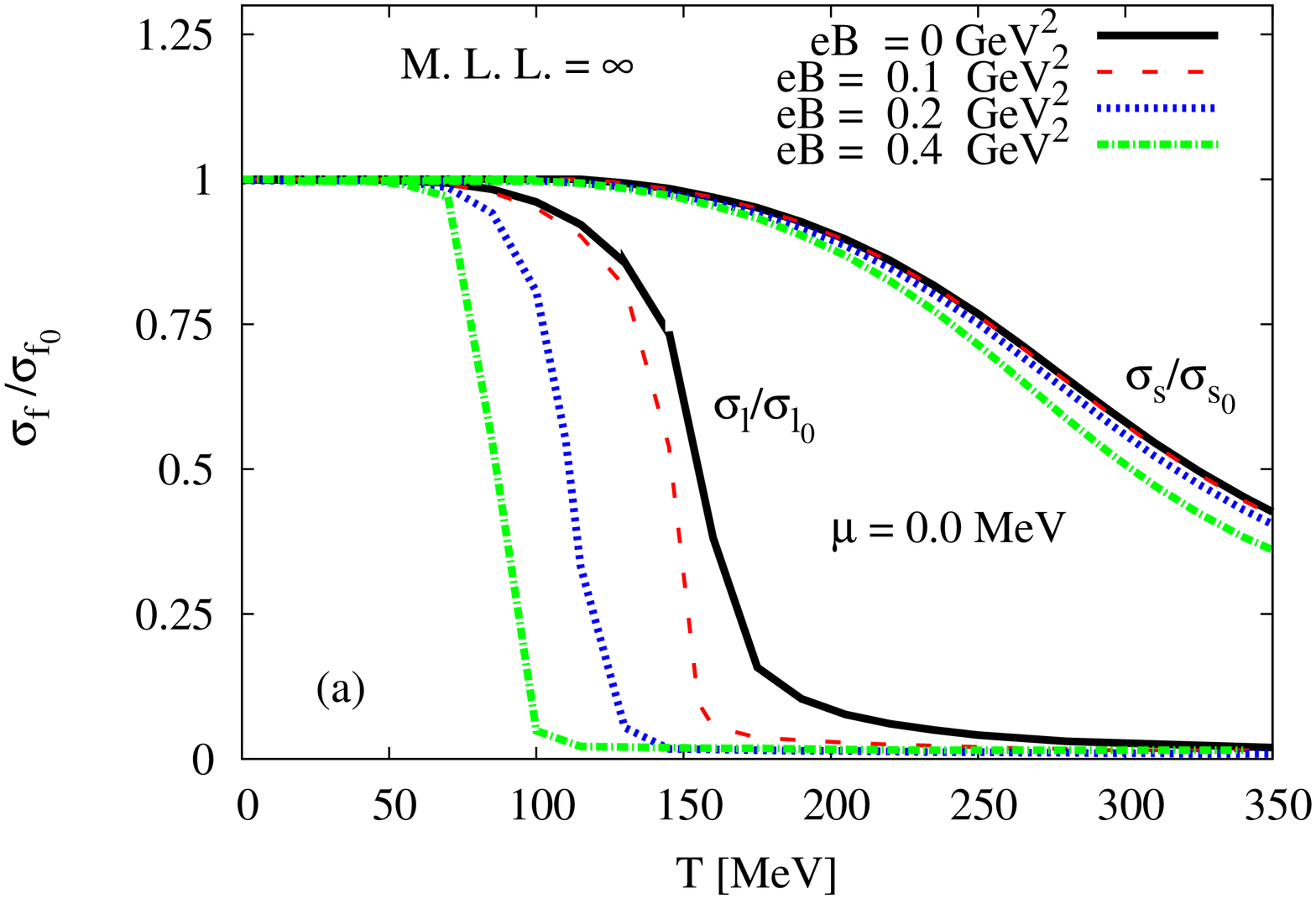}
\includegraphics[width=5cm,angle=-90]{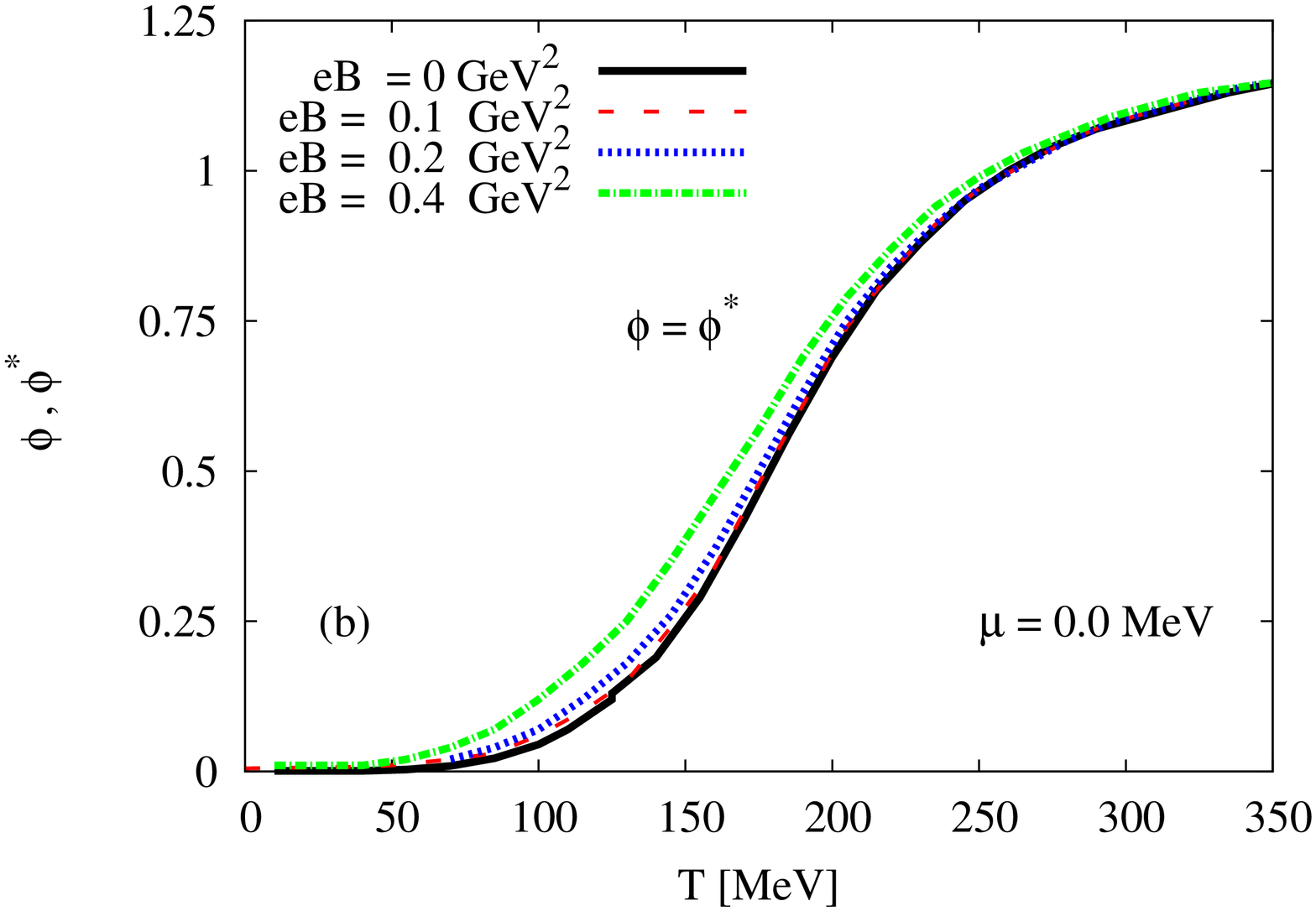}
\caption{Left--hand panel: the chiral condensates normalized to the vacuum value are given as functions of temperature. Right--hand panel: the expectation values of Polyakov loop parameters, $\phi$  and $\phi^*$, are calculated in dependence on $T$, at $eB=0$ (solid curves), $0.1$ (dashed curves), $0.2$ (dotted curves) and $0.4~$GeV$^2$ (dot--dashed curves) and vanishing baryon chemical potential. \label{fig:sbtrc2}
}}
\end{figure*}

Figure \ref{fig:sbtrc2} depicts the temperature dependence of normalized chiral condensates (a) and deconfinement order parameters (b), at $eB=0$ (solid curves), $0.1$ (dashed curves), $0.2$ (dotted curves) and $0.4~$GeV$^2$ (dot--dashed curves) and vanishing baryon chemical potential. The left--hand panel (a) depicts the same as  the left-hand panel of Fig. \ref{fig:sbtrc1} but for different values of the magnetic field, at fixed values of baryon chemical potential. We notice that the chiral critical temperature deceases and the crossover phase--transition becomes sharper with increasing magnetic field. This can be interpreted due to the maximum occupation of the Landau levels ($\nu_{max}\rightarrow \infty$). We conclude that the phase transition seems to be of first order, whenever the chiral condensate passes through a meta--stable phase, in which light quarks become massless and move freely.  

In right--hand panel (b), the temperature dependence of the deconfinement phase--transition is depicted, at $\mu=0$. $\phi=\phi^*$, at $eB=0$ (solid curves), but having different values, ar $0.1$ (dashed curves), $0.2$ (dotted curves) and $0.4~$GeV$^2$ (dot--dashed curves). It is obvious that the deconfinement critical temperature, $T_{\phi}$, slightly decreases as the magnetic field increases.

It would be noticed that if one compares the curve for $eB=0.1~$GeV$^2$ with the curve of $eB=10 m_\pi^2\simeq0.2~$GeV$^2$, which would be nearly similar values, one would find a slight difference. This is an artifact. It comes from the effect of the occupation Landau levels. The Landau levels and the  quantization number are determined for the medium parameters such as $T$, $\mu$, $eB$.

\begin{figure}[htb]
\centering{
\includegraphics[width=4.5cm,angle=-90]{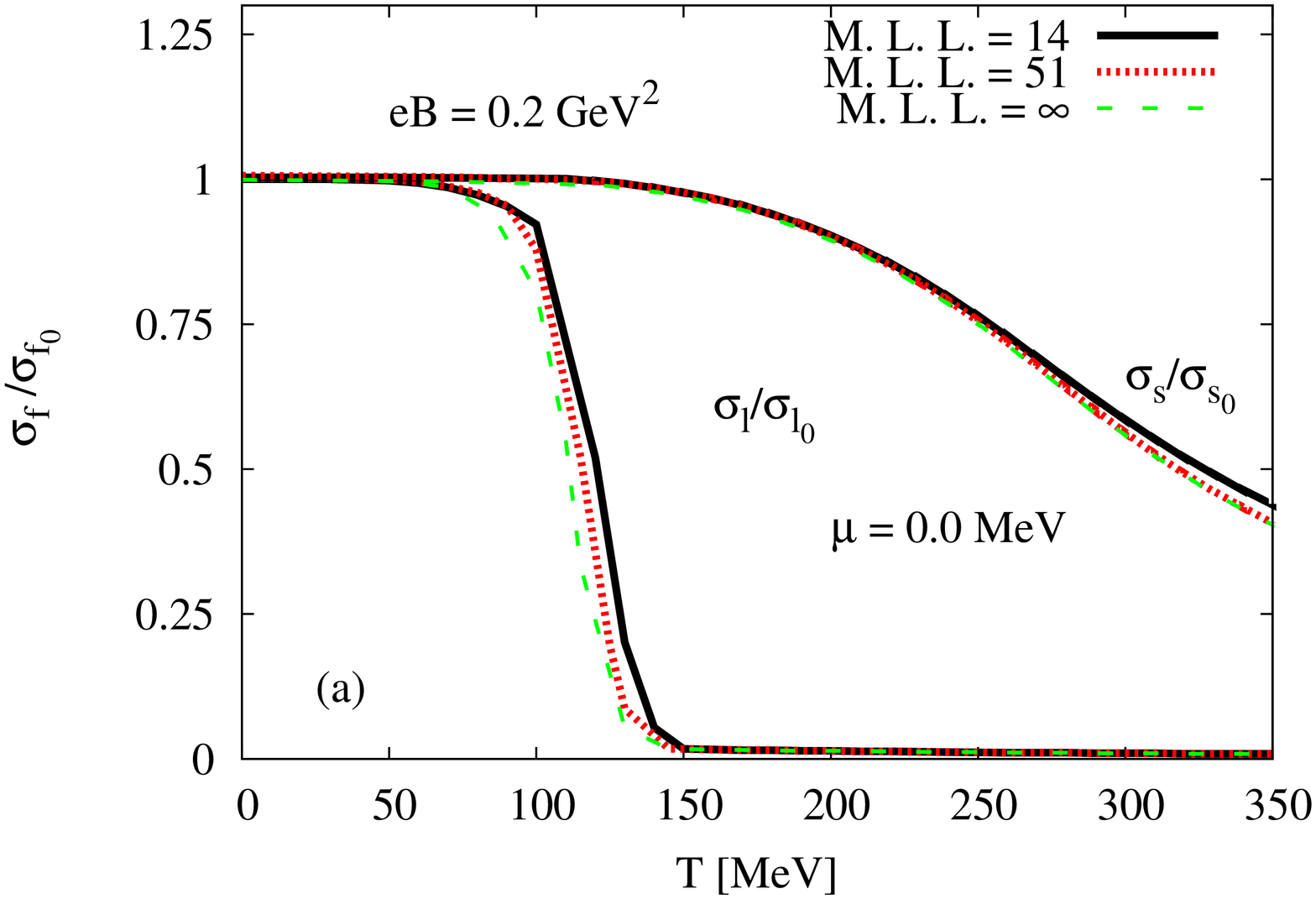}
\caption{The normalized chiral condensates are given as functions of temperatures, at $eB=0.2~$GeV$^2$. Different MLL are taken into considration. \label{fig:sbtrc3}
}}
\end{figure}

As discussed, PLSM is well--suited to study the chiral limit. The inclusion of magnetic field can be achieved by modifying the dispersion relation of quarks and antiquarks, Eq. (\ref{PloykovPLSM}). In doing this, the dimension of the momentum--space should be reduced from three to one and scaled via quark charge and magnetic field. This process is known as magnetic catalysis effect of dimension reduction \cite{Shovkovy:2012zn}. Furthermore, the introduction of the magnetic field requires suitable implementation of the Landau quantization, section \ref{sec:llquant}. 

Figure \ref{fig:sbtrc3} shows the effects of the occupation number of the Landau levels on the temperature dependence of the chiral condensate, $\sigma_l$ and $\sigma_s$, at a finite magnetic field $eB=0.2~$GeV and a vanishing baryon chemical potential. We observe that the change in the Landau levels is relatively significant only in crossover phase--transition  region and seems to disappear otherwise. At $MLL=14$ (solid curves), $51$ (dotted curves) and $\infty$ (dashed curves), the normalized chiral condensates for light and strange quarks are analyzed as functions of temperatures, at finite magnetic field and vanishing chemical potential. We conclude that the increase in the Landau levels very slightly sharpens the phase transition or the crossover and decreases the critical temperature $T_{\chi}$.

%
%
\begin{figure*}[htb]
\centering{
\includegraphics[width=5cm,angle=-90]{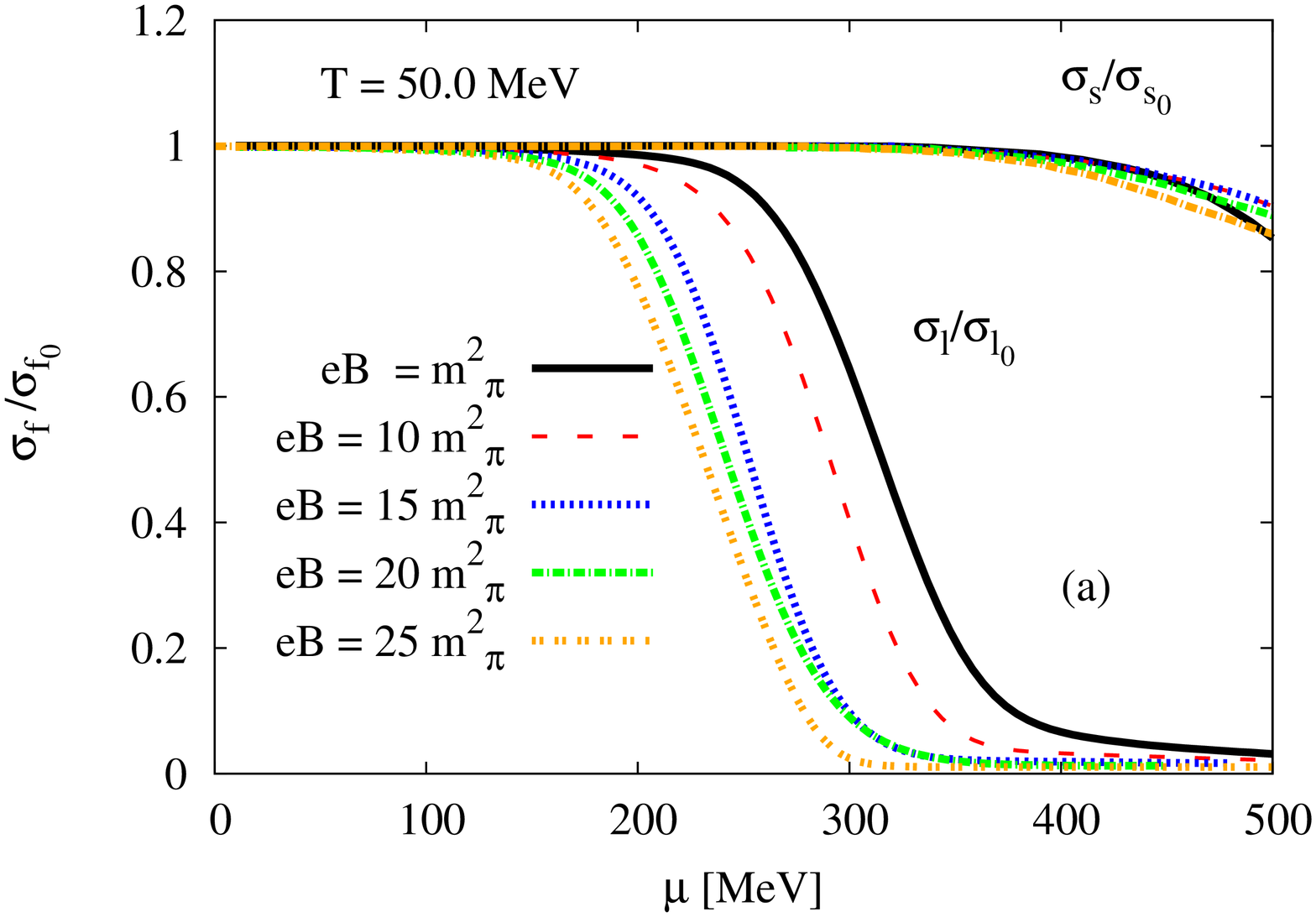}
\includegraphics[width=5cm,angle=-90]{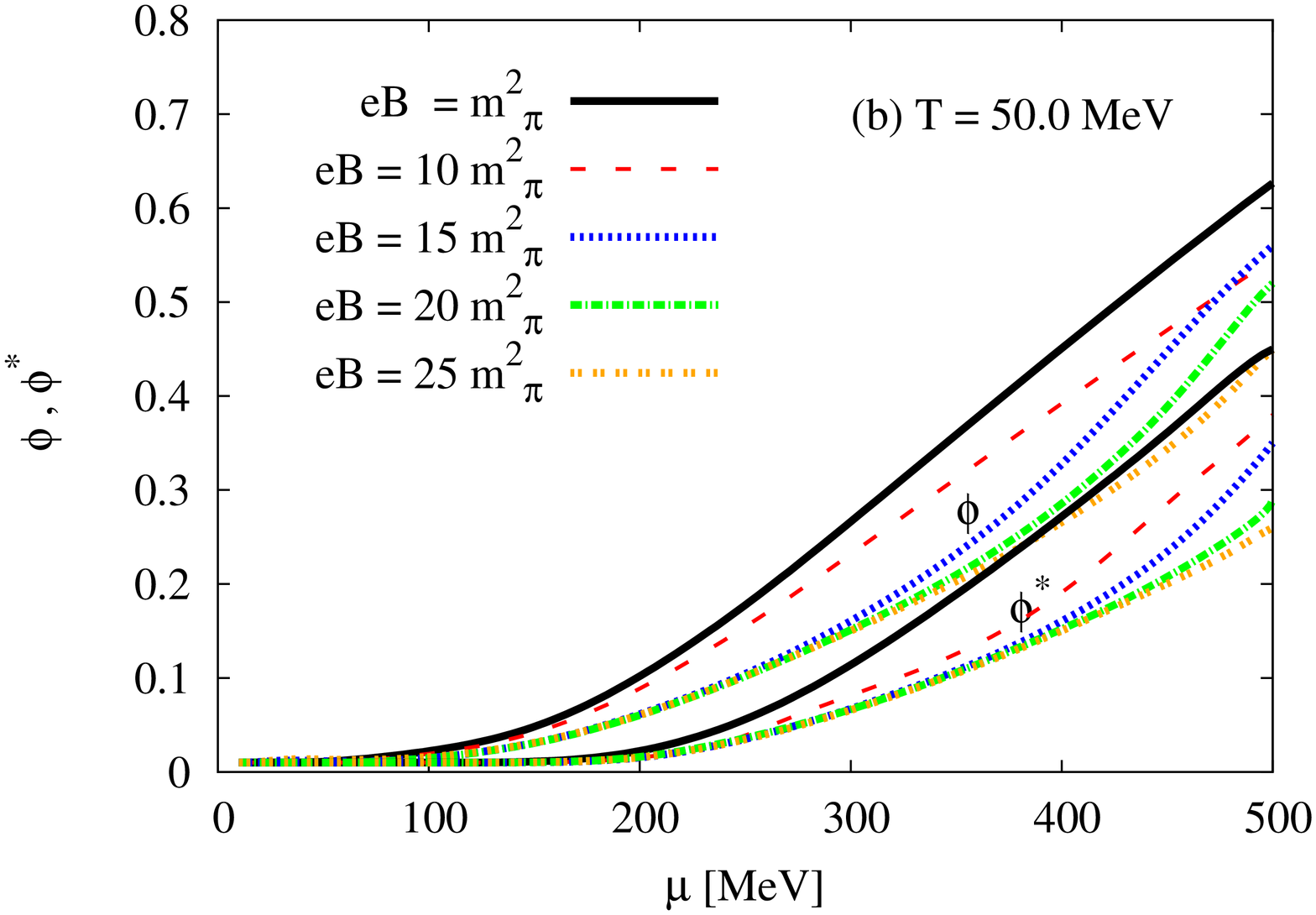}\\
\includegraphics[width=5cm,angle=-90]{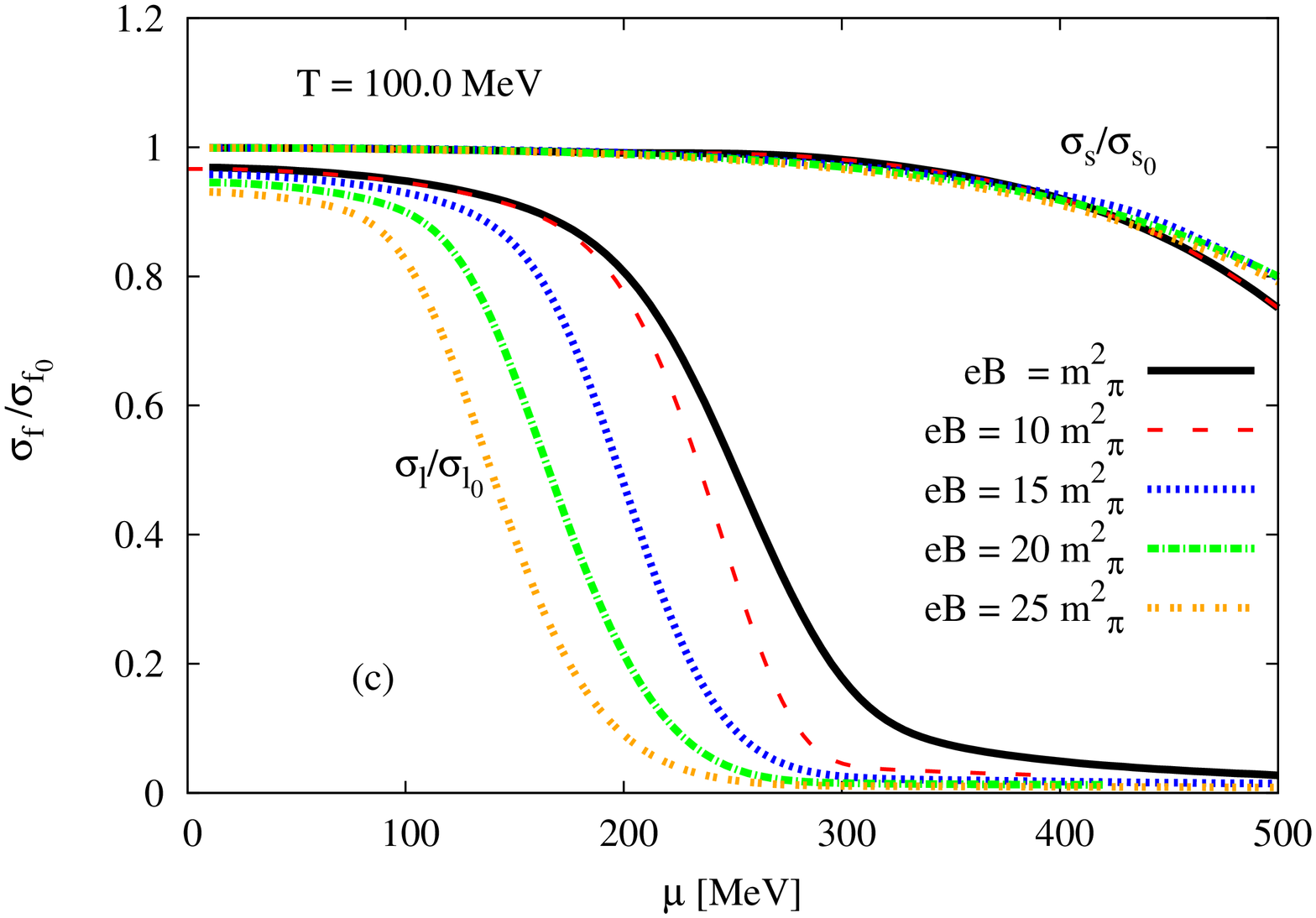}
\includegraphics[width=5cm,angle=-90]{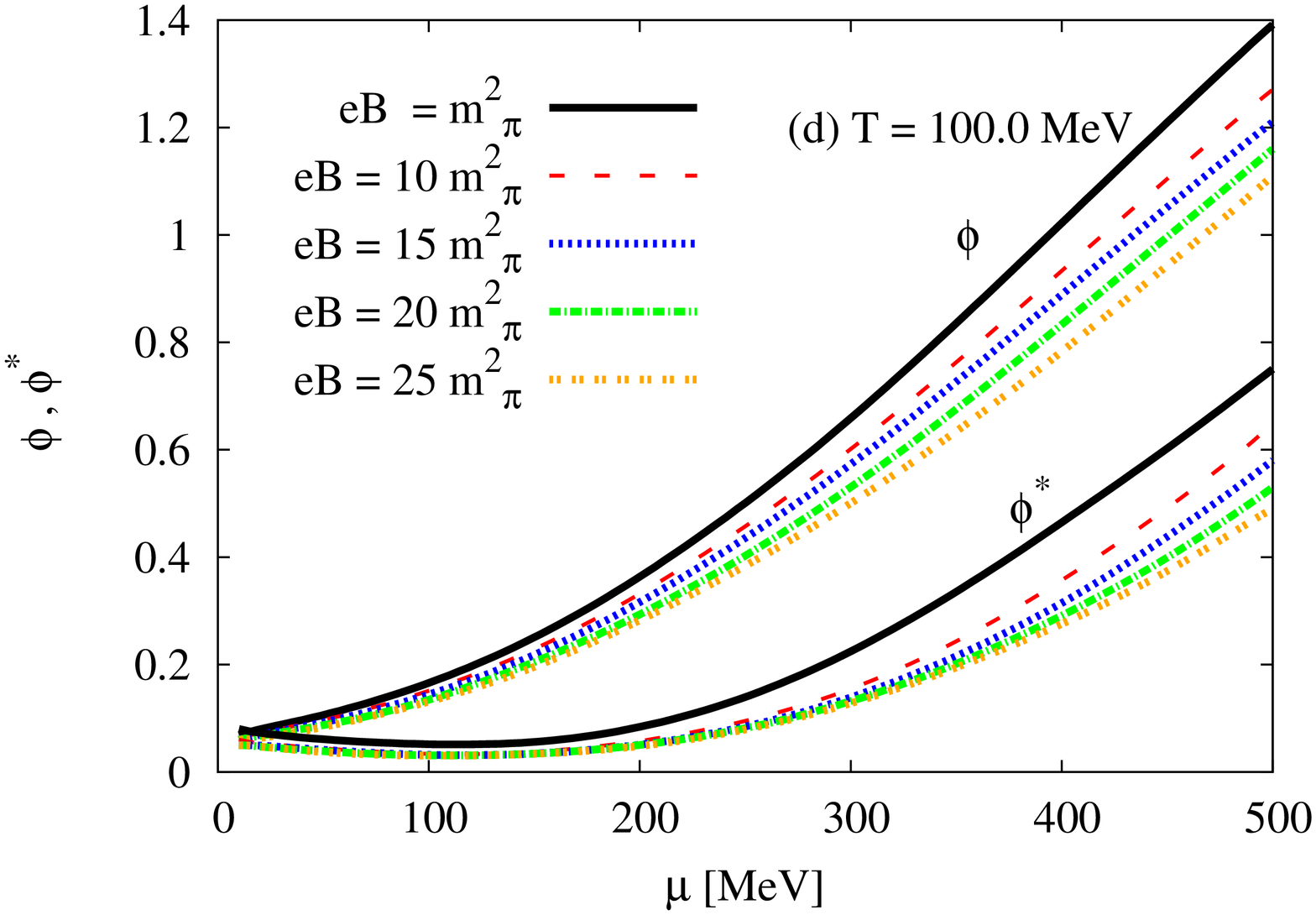}
\caption{Left--hand panels (a) and (c) show the chiral condensates, $\sigma_l$ and $\sigma_s$, normalized to the vacuum value as functions of the baryon chemical potentials, at $eB=1$ (solid curves) $10$ (dashed curves), $15$ (dotted curves), $20$ (dot--dashed curves), and $25\,m^2_{\pi}~$ (double dotted curves). Right--hand panels (b) and (d) give the expectation values of the Polyakov loop parameters, $\phi$ and $\phi^*$, as functions of baryon chemical potentials, at finite temperatures. Top panel shows the results, at $T=50\,$MeV, while the bottom panel at $T=100\,$MeV. \label{fig:sbtrcMu}
}}
\end{figure*}

Furthermore, the temperature dependence of the chiral condensates, $\sigma_l$ and $\sigma_s$, and the deconfinement order--parameters, $\phi$ and $\phi^*$, are estimated, at different values of magnetic fields ($eB$) and baryon chemical potential ($\mu$). Fig. \ref{fig:sbtrcMu} shows the deconfinement order--parameters as functions of baryon chemical potentials, at $T=50~$MeV (upper--panel) and $T=100~$MeV (bottom--panel), from which we notice that the magnetic effect is very obvious. In left--hand panels (a) and (c), the chiral condensates are given in dependence on magnetic fields; $eB=1$ (solid curves) $10$ (dashed curves), $15$ (dotted curves), $20$ (dot--dashed curves), and $25\,m^2_{\pi}~$ (double dotted curves). Firstly, we conclude that the temperature causes a rapid decrease in the chiral condensates around the chiral phase--transition similar to what was observed in a previous study from PLSM without magnetic field \cite{Tawfik:2014gga}. Secondly, a small sudden drop in the chiral condensates referring to first--order phase--transition takes place. There is a gap difference between light and strange chiral condensates, at very high density. This could be understood because of the inclusion of the anomaly term in Eq. (\ref{pure:meson}), where the fit parameters are accordingly modified \cite{Schaefer:2008hk,Tawfik:2014gga}. This was conjectured as an evident on numerical estimation of the chiral condensates. The magnetic field seems to have a sudden drop in the chiral phase--structure. This causes an increase in the critical temperatures. We can draw a conclusion that increasing magnetic field tends to sharpen the phase transition and to accelerate the formation of metastable phase.

The right--hand panels (b) and (d) give the deconfinement phase--transition in dependence on magnetic fields; $eB=1$ (solid curves),  $10$ (dashed curves), $15$ (dotted curves), $20$ (dot--dashed curves), and $25\,m^2_{\pi}$ (double dotted curves). The increase in temperature increases the deconfinement phase--transition to larger baryon chemical potential. The thermal and magnetic effects of the hadronic medium on the evolution of Polyakov loop parameters seem to be very smooth. The slope of $\phi$ and $\phi^*$ to the baryon chemical potential can approximately be estimated. It is obvious that the magnetic field decreases these slopes (increases the critical temperature), while the temperature increases them (decreases the critical temperature).

\subsection{Thermodynamics}

We start this analysis with a brief introduction to the basic quantities of thermodynamics, at non--vanishing magnetic field strength. For a statistical system in equilibrium with volume $V$, temperature $T$ and chemical potential $\mu$, the grand--canonical density operator, $\hat{\rho}$, the grand--canonical partition function, $\mathcal{Z}(T, V, \mu)$, and the grand potential, $\Omega(T, V, \mu)$ can be introduced in the natural units $\kappa_B=\hbar=c=1$. The pressure as a function of finite magnetic field $eB$ reads 
\bea
P(T, \mu, eB) = - \Omega (T, \mu, eB),
\eea
which enables us to characterize the phenomenology to the strong interacting QCD matter, at finite magnetic fields, as the case in HIC, including magnetization, magnetic susceptibility and permeability. 
    
Then, the free energy density can be written as \cite{Aad:2012ew}
\bea
\mathbf{f} &=& \epsilon - T s = \epsilon^{\mbox{tot}} - \epsilon^{\mbox{field}} - T s   \\ 
&=& \epsilon^{\mbox{tot}} - T s - eB \; \mathcal{M}, \label{totalfree}
\eea
where $\mathcal{F} = -T \log \cdot \mathcal{Z}$, the total energy density $\epsilon^{\mbox{tot}} =\epsilon + \epsilon^{\mbox{field}}$, which in turn is divided into two terms; one for the energy density of the medium $\epsilon$ and another one of the magnetic field $\epsilon^{\mbox{field}} = eB \mathcal{M}$. 

We notice that the partition function in vanishing magnetic field is given by an integral over six--dimensional phase--space. The dispersion relations follow the Lorentz invariance principle. But, in finite magnetic field, the integral dimensional is reduced and simultaneously accompanied by a considerable modification in the dispersion relation. 

The velocity of a test particle with momentum $\partial \epsilon^{\mbox{tot}}/\partial P$ in finite magnetic field $B$ can be expressed as
\bea
v_p &=&  c \left[\frac{c\, p}{c\, p + 2 |q_f|(\kappa+\frac{1}{2}-\frac{\sigma}{2}) B}\right], 
\eea
where $\sigma=\pm S/2$. Then, the causality is guaranteed for $v_p$ not--exceeding the speed of light $c$, i.e. as long as the $B$--term is finite positive, which should be quantitatively estimated as a function of temperature and magnetic field. 
From Eq. (\ref{totalfree}), the entropy density and the magnetization, for instance, could be derived as 
\bea
s = - \frac{1}{V} \frac{\partial \mathcal{F}}{\partial T} , \quad \quad \quad \mathcal{M} = - \frac{1}{V} \frac{\partial \mathcal{F}}{\partial (eB)}. \label{Magnet_entropy}
\eea

Other thermodynamic quantities, such as, pressure can be derived, as well. Because the magnetic field is conjectured to mark a preferred direction, $P_i$ might be different along the geometry effect of the magnetic field. As $V= L_x L_y L_z$ and the magnetic field is distributed along $z$ direction, we can distinguish between two different systems.
\begin{itemize} \itemsep0em
\item $B$--scheme, in which the magnetic field remains constant in direction, results in an isotropic pressure 
\bea
P_x = P_y = P_z = - \mathbf{f}
\eea
\item $\varphi$--scheme, which sets up magnetic flux $\varphi = eB \cdot L_x L_y$, remains constant and results in an anisotropic pressure 
\bea
P_x = P_y = P_z - eB \mathcal{M}
\eea 
\end{itemize}
Accordingly, the thermodynamic quantities should be modified. For instance, in $\Phi$--scheme, the trace anomaly (the interaction measure) reads
\bea
I &=& \epsilon - 3p_z + 2  eB \cdot \mathcal{M}. \label{InteractionMeasure}
\eea  
The speed of sound squared, at constant entropy, is given as \cite{Tawfik:2012ty},
\bea
c_s^2 &=& \left(\frac{\partial p}{\partial \epsilon}\right)_s =\frac{\partial p}{\partial T}/ \frac{\partial \epsilon} {\partial T} 
= \frac{s}{c_v}, \label{eq:cs2}
\eea
where the specific heat, $c_v$, gives the thermal rate change of the energy density, at constant volume. In finite magnetic field, Stefan--Boltzmann (SB) limits can be deduced from lowest--order perturbation theory \cite{Bali:2014kia}
\bea
T\, \log \, \mathcal{Z}(V,T,\mu, eB) &=& \frac{19\, V \pi^2}{36} T^4 \nn\\
&+& b_1^{\mbox{free}} (eB)^2\, V \, \log\left(\frac{T}{\Lambda_H}\right)+\cdots, \hspace*{6mm}
\eea

\begin{figure*}[htb]
\centering{
\includegraphics[width=5.cm,angle=-90]{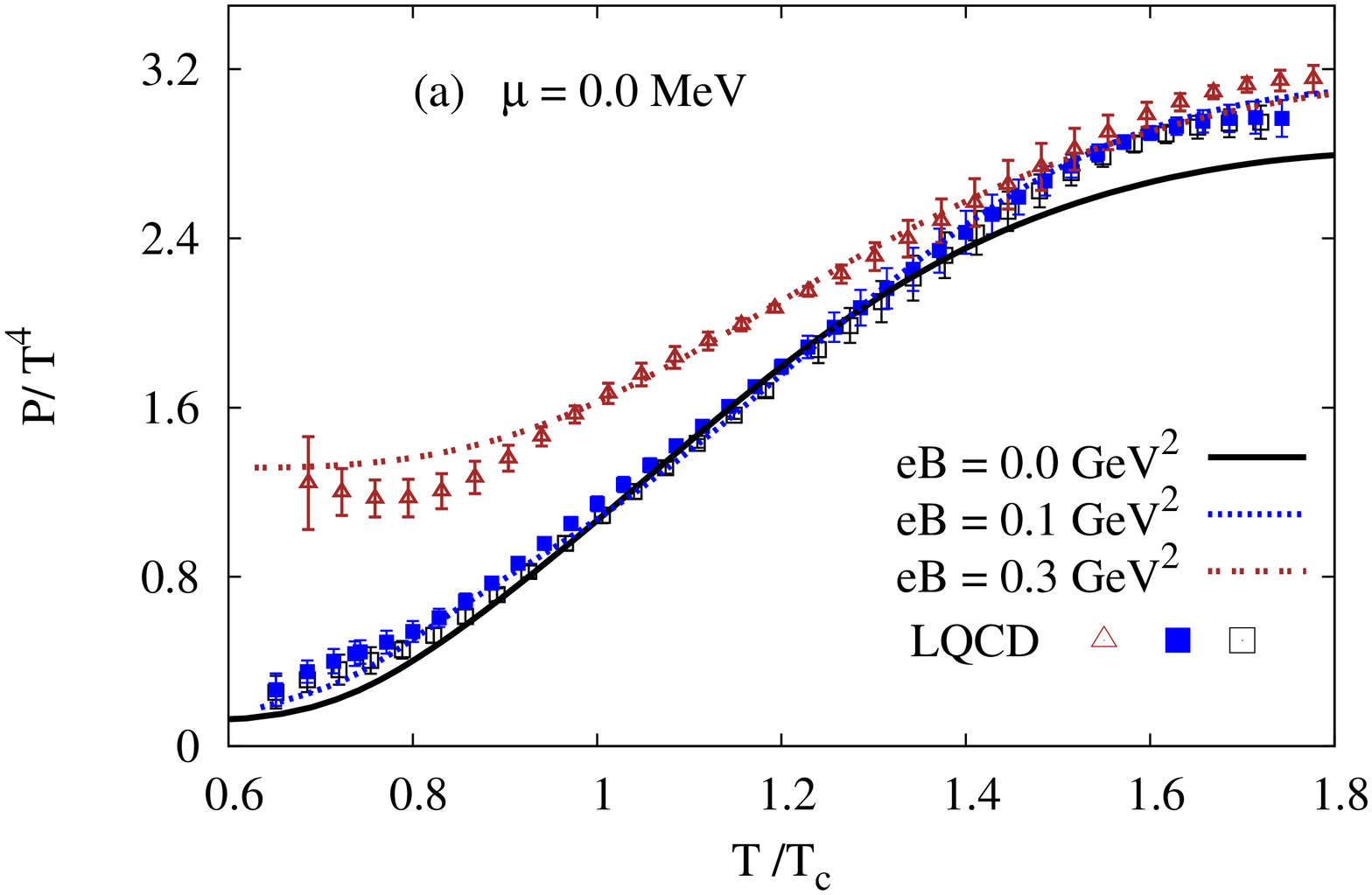}
\includegraphics[width=5.cm,angle=-90]{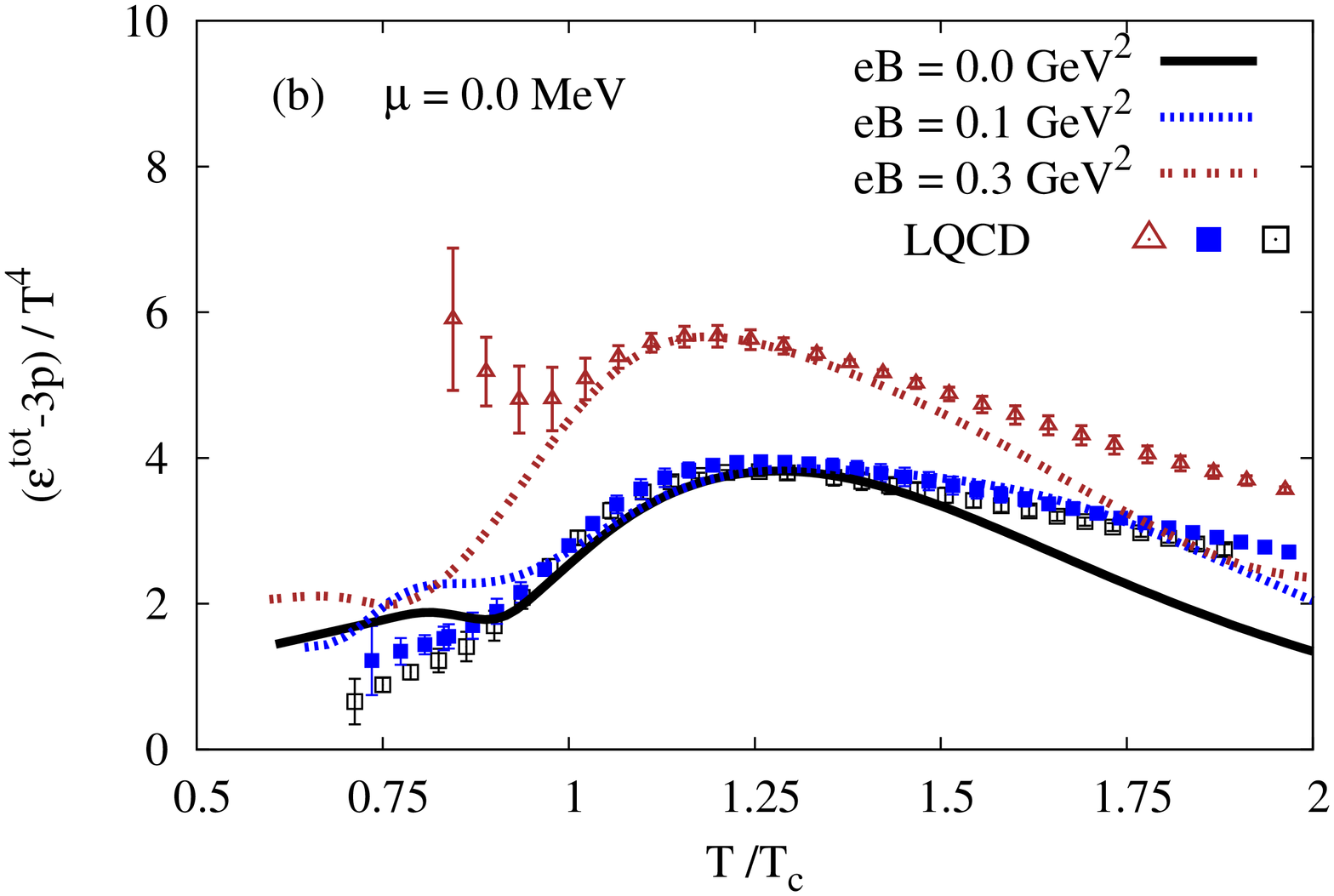}
\caption{Left--hand panel (a): the normalized pressure $p/T^4$ versus $T/T_{c}$ is calculated in PLSM, at finite magnetic fields $eB=0.0$ (solid curve), $eB=0.1$ (dotted curve) and $eB=0.3~$GeV$^2$ (double--dotted curve) and compared with recent lattice QCD (open square), (close square) and (open triangle), respectively \cite{Bali:2014kia}. Right--hand panel (b): The same as in the left--hand panel but for the normalized trace--anomaly $(\epsilon^{\mbox{tot}}-3P)/T^4$. \label{fig:prssTraceeB}}}
\end{figure*}

Figure \ref{fig:prssTraceeB} depicts the normalized pressure $p/T^{4}$ (left--hand panel) and the normalized trace--anomaly $(\epsilon^{\mbox{tot}}-3P)/T^4$ (right--hand panel) as functions of temperatures, at vanishing baryon chemical potential but finite values of magnetic fields $eB=0.0$ (solid curve), $eB=0.1$ (dotted curve) and $eB=0.3~$GeV$^2$ (double--dotted curve). The results are compared with recent lattice QCD \cite{Bali:2014kia} (open square), (close square) and (open triangle), respectively. It is obvious that the pressure increases with increasing magnetic field, especially, at low temperatures. At high temperatures, $p/T^4$ is limited to the SB limits, which apparently negligibly are affected by the magnetic field. 

The right--hand panel of Fig. \ref{fig:prssTraceeB} presents the modified normalized trace--anomaly, Eq. (\ref{InteractionMeasure}), as a function of temperature and magnetic field strengths, $eB=0.0$ (solid curve), $eB=0.1$ (dotted curve) and $eB=0.3~$GeV$^2$ (double--dotted curve). These are also compared with recent lattice QCD \cite{Bali:2014kia} (open square), (close square) and (open triangle), respectively. 

The  normalized entropy density, $s/T^3$, which is derived from $p_z$ with respect to $T$,  vanishes at $T=0$. This may be understood from the fact that the vacuum contribution is conjectured to be a pure quantum effect. This emerges from the interaction of virtual quarks with the external field. Thus, it likely doesn't contribute to entropy \cite{Bali:2014kia}. We also notice that, at $T>0$, the magnetic field changes the thermal distributions and is expected to modify entropy, 
\bea
s = \frac{\epsilon + p_z}{T}.
\eea 
We find that near the $T_c$--regime, $s/T^3$ excellently agrees with the lattice QCD calculations. This might not be also the case, at higher temperatures.

\begin{figure*}[htb]
\centering{
\includegraphics[width=5cm,angle=-90]{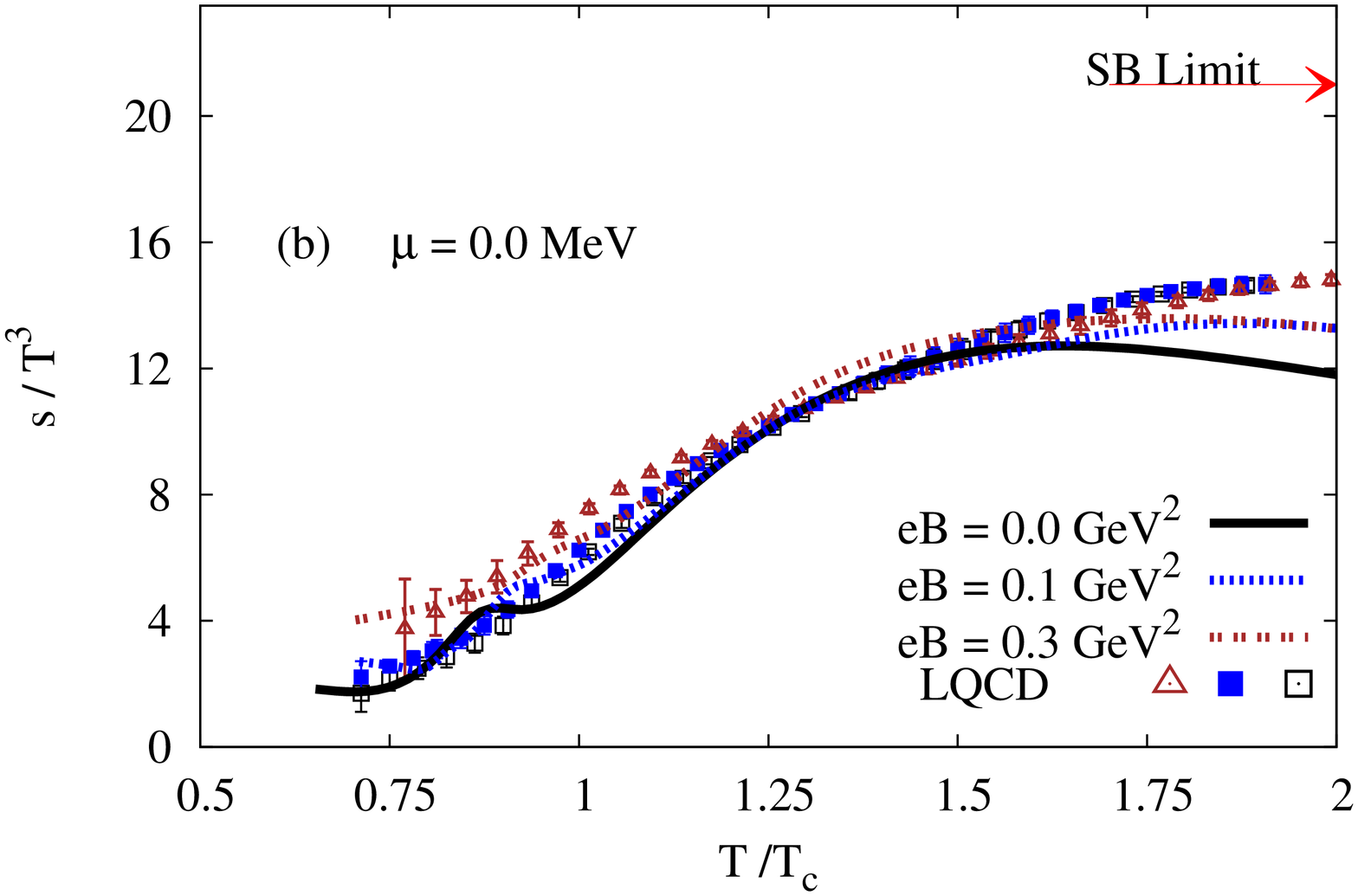}
\includegraphics[width=5cm,angle=-90]{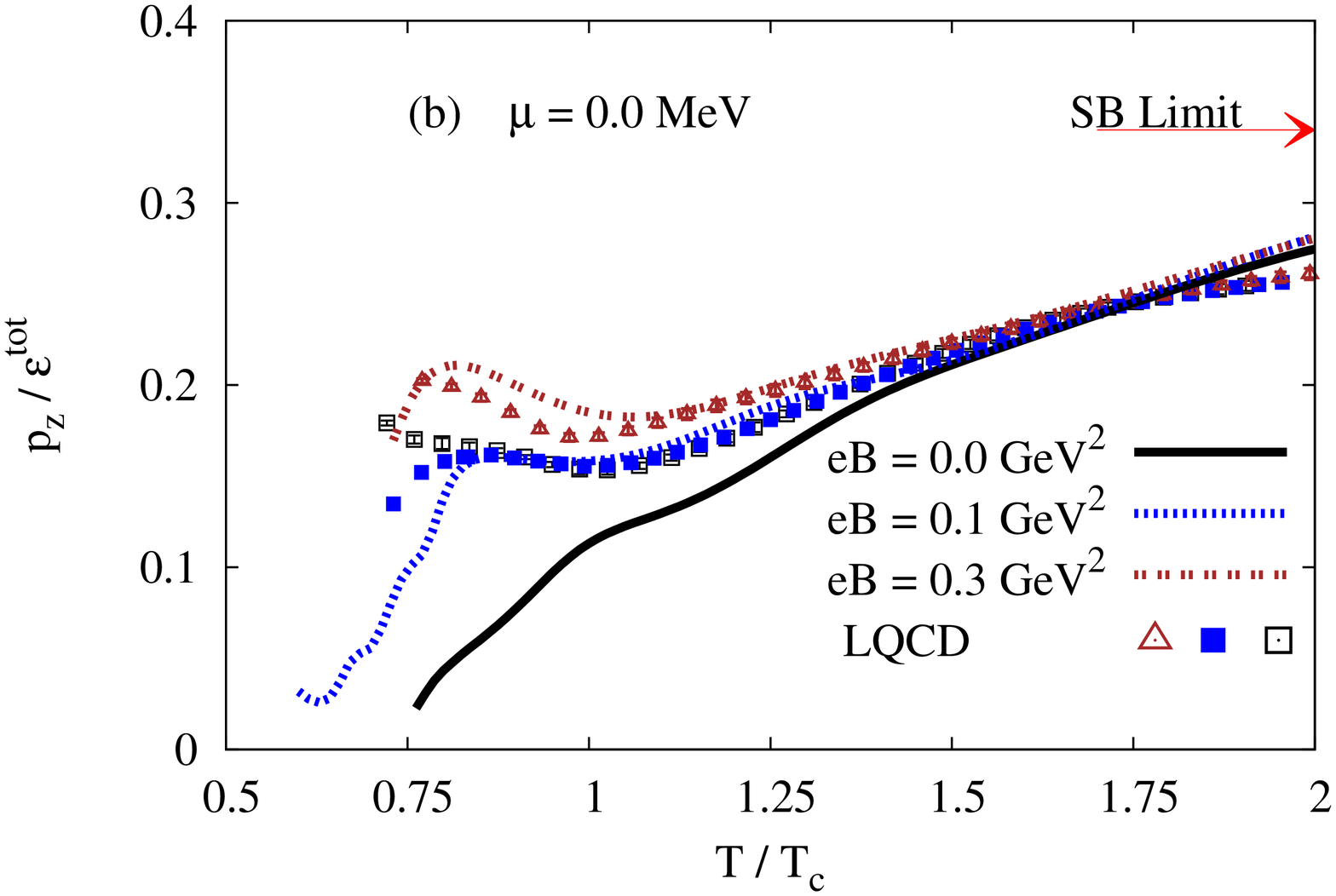}
\caption{Left--hand panel (a): the normalized entropy $s/T^3$ versus $T/T_{c}$ calculated in PLSM, at finite magnetic fields, $eB=0.0$ (solid curve), $eB=0.1$ (dotted curve) and $eB=0.3~$GeV$^2$ (double--dotted curve). The results are compared with recent lattice QCD (open square), (close square) and (open triangle), respectively \cite{Bali:2014kia}. Right--hand panel (b) shows the same as in the left--hand panel but here $P\epsilon^{\mbox{tot}}$.\label{fig:entropyspeedEB}}}
\end{figure*}

Figure \ref{fig:entropyspeedEB} presents $s/T^3$ and the equation of state $\epsilon (p)$ included in $\epsilon^{\mbox{tot}}$ in dependence on $T$, at $\mu=0~$MeV and the same values of the magnetic field as depicted in Fig. \ref{fig:prssTraceeB}, namely $eB=0.0$ (solid curve), $eB=0.1$ (dotted curve) and  $eB=0.3~$GeV$^2$ (double--dotted curve). The corresponding curves and lattice points are the same as in Fig. \ref{fig:prssTraceeB}. We find a reasonable agreement with the lattice QCD simulations The phase transition seems to smoothly take place. The temperature dependence continues even above $T_c$. $s/T^3$ keeps its increase with increasing $T/T_c$ so that it becomes slightly lower than the lattice results.

The right--hand panel of Fig.  \ref{fig:entropyspeedEB} (b) presents the PLSM calculations for $p/\epsilon$, at $eB=0.0$ (solid curve), $eB=0.1$ (dotted curve) and $eB=0.3~$GeV$^2$ (double--dotted curve). We also compare with recent lattice QCD \cite{Bali:2014kia} (open square), (close square) and (open triangle), respectively. A fair agreement is also obtained, especially at $eB=0.1$, at low temperature. It is obvious that such an agreement could be improved with increasing magnetic fields.

Now, we are able to estimate some fundamental properties of the strongly interacting QCD matter in finite magnetic field such as the magnetization, the magnetic susceptibility, and the permeability. The response of QCD matter to an external magnetic field can be estimated from the free energy density. The magnetic susceptibility with proper renormalization has been introduced in litrature \cite{Kamikado:2014bua}. The magnetic permeability measures the ability of the QCD matter to generate magnetic field or the ability to store magnetic potential energy, which is proportionally constant for the magnetic flux. The magnetic flux - in turn - is produced from influences of the magnetic field. The magnetic permeability can be calculated along the magnetic field on the transverse direction to momentum space $p_z$. 

In thermal, dense and magnetic medium, the partition function $\ln\, \mathcal{Z}$ is to be properly modified, from which the magnetization can be derived, Eq. (\ref{Magnet_entropy}).  The sign of magnetization defines an essential magnetic property, namely whether QCD matter is {\it para}-- or {\it dia}--magnetic, i.e. $M>0$ (bara) or $M<0$ (dia). 
\begin{itemize} \itemsep0em
\item For {\it dia}--magnetized QCD matter, the color charges align oppositely to the direction of the magnetic field and produce an induced current, which spreads as small loops attempting to cancel the effects of the applied magnetic field, and 
\item For {\it para}--magnetized QCD matter, most color charges align towards the direction of the magnetic field. 
\end{itemize}

\begin{figure}[htb]
\centering{
\includegraphics[width=5.5cm,angle=-90]{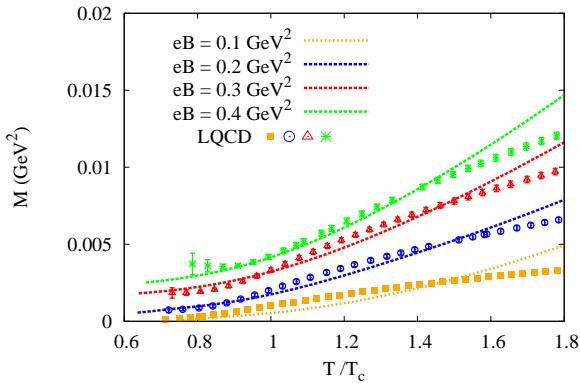}
\caption{The magnetization $\mathcal{M}$ is calculated as a function of $T$, at $eB = 0.0-0.4$ GeV$^2$ and compared with recent lattice QCD simulations (symbols) \cite{Bali:2014kia}. \label{fig:magnetization}}}
\end{figure}

As discussed, the magnetization greatly affects the thermodynamic properties of the QCD matter, as the magnetization measures the response of the system of interest to finite magnetic field. The latter is likely extremly generated in HIC, at least in very short time intervals. Aslo, because of the relativistic, off--center motion of spectators, i.e. peripheral collisions, the rapid motion of electric charges generates magnetic field perpendicular to plane of both motion direction and the electric field. Also, because of local imbalance in the momenta carried by the colliding nucleons in peripheral and central collisions. This local imbalance leads to an angular momentum and thus a magnetic field \cite{Tawfik:2016cot}. As discussed, such a magnetic field is typically very huge, ${\cal O}(m_{\pi}^2)$. It largely exceeds the detector's magnet field.

The magnetization, $\mathcal{M}$, can be derived from Eq. (\ref{potential}). The values obtained can be given in GeV$^2$ in the natural units. The sign of $\mathcal{M}$ refers to {\it para}-- or {\it dia}--magnetic QCD matter. If $\mathcal{M}>0$ or $\mathcal{M}<0$, the QCD matter is either {\it para}-- or {\it dia}--magnetized, respectively. 

In Fig. \ref{fig:magnetization}, $\mathcal{M}$ is given as a function of $T$, at $eB= 0.1$ (dotted), $0.2$ (dashed), $0.3$ (double--dotted) and $0.4$ GeV$^{2}$ (dash--dotted curve) at vanishing $\mu$. The results are compared with recent lattice QCD \cite{Bali:2014kia} at $eB=0.1$ (closed square), $0.2$ (circle), $0.3$ (triangle) and $0.4$ GeV$^{2}$ (astride). It is obvious that $M>0$ and increases as the magnetic field increases. This result indicates that paramagnetic properties of the QCD matter. At temperatures below the critical value, the PLSM results resemble the lattice data in an excellent way. At temperatures characterizing QGP (higher than the critical temperature), the PLSM curve becomes larger than the lattice data, especially at very high temperatures. In this range of temperatures, the colorless hadrons are conjectured to deconfine into colored quarks and gluons. The discrepancies suggest that the corresponding dof are not sufficient enough to achieve a good agreement, especially at very high temperature. Furthermore, we notice that the PLSM calculations give an evident on paramagnetic features of the hot QCD matter.

\begin{figure*}[htb]
\centering{
\includegraphics[width= 8.cm,angle=0]{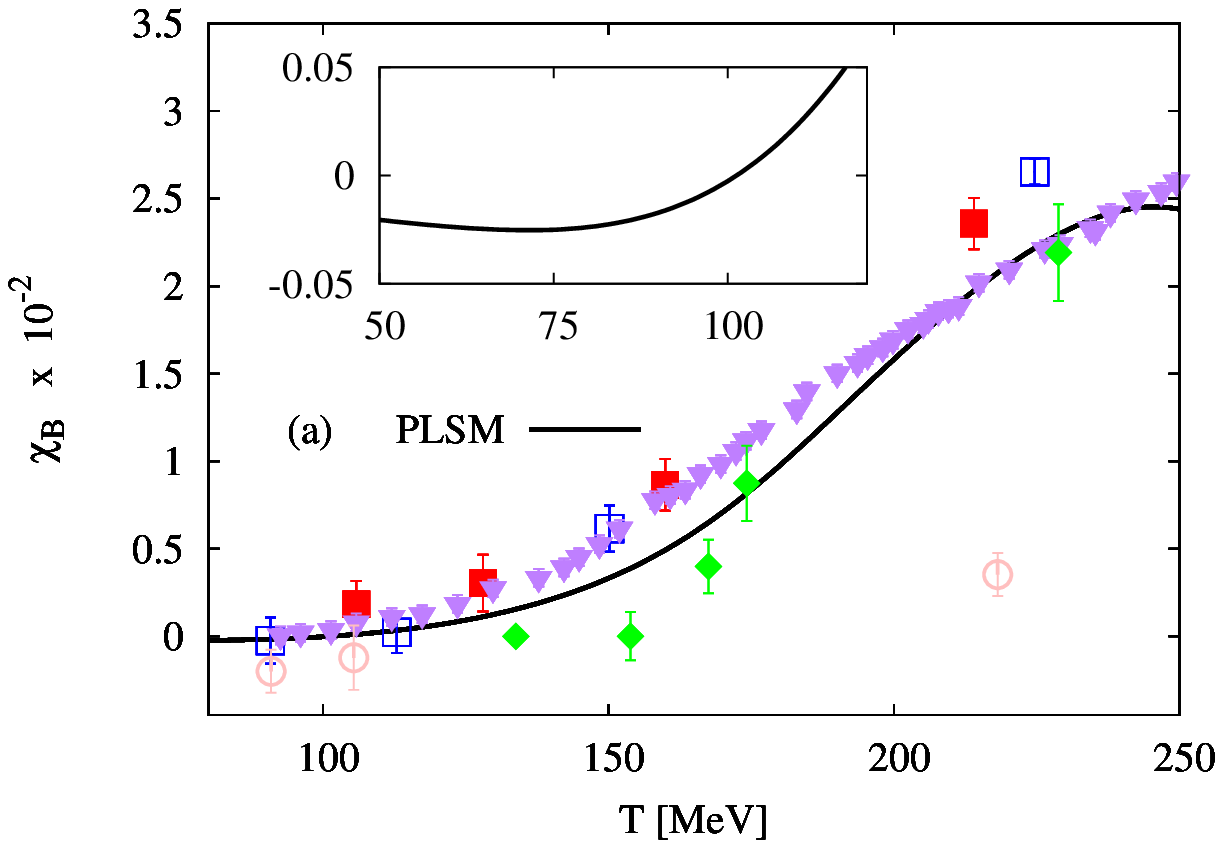}
\includegraphics[width= 8.cm,angle=0]{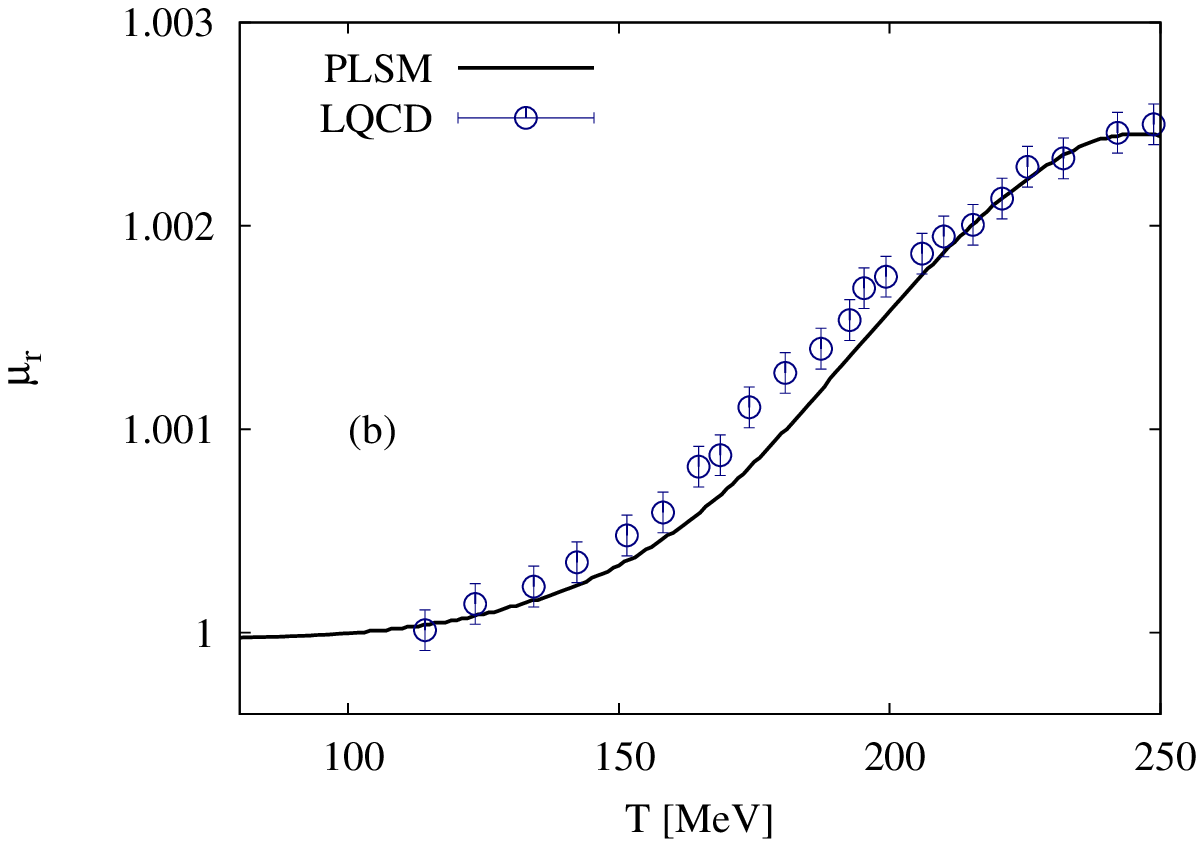}
\caption{Left--hand panel (a): the magnetic susceptibility $\chi_B$ versus $T/T_{c}$ calculated in PLSM, at $eB = 0~$GeV$^2$. Right--hand panel (b): The same as in the left--hand panel but for the magnetic permeability. \label{fig:susceptibility}}}
\end{figure*}

Also, the magnetic susceptibility and permeability reflect the magnetic response of the hot QCD matter. In other words, the response of QCD matter to the magnetic field can be determined by the slope of $\mathcal{M}$ with respect to the magnetic field. The second derivative of the free energy density with respect to the finite magnetic field is the magnetic susceptibility, 
\bea
\chi_B = - \frac{1}{V} \frac{\partial^2 \mathcal{F}}{\partial (eB)^2}\vert_{eB=0}.
\eea
In response to the magnetic field, the magnetic susceptibility is a dimensionless proportionality parameter indicating the degree of magnetization of the QCD matter. 

Furthermore, the relative magnetic permeability, $\mu _{r}$, normalized to the vacuum magnetic permeability $\mu_0$ can be translated as the magnetic effect in thermal QCD medium. This can be determined by different methods. With a direct relation to the magnetic susceptibility, we have
\begin{equation}
\mu_{r} = 1 + \chi_B.
\end{equation} 
As shall be introduced, this relation agrees well with the lattice QCD simulations, in which the magnetic permeability is expressed in terms of the magnetic susceptibility,
 \begin{equation}
\mu_B \equiv \frac{B^{ind}}{B^{ext}} = \frac{1}{1 - 4 \pi \alpha_{m} \cdot \chi _B}, \label{permeability}
\end{equation}
where $\alpha_{m}=e^2 /4\pi$ is the fine structure constant. This expression distinguishes between dimensionless proportionality constants, namely external $B^{ext}$ and induced magnetic field $B^{ind}$. We highlight that the higher--order permeability seems to be limited by the magnetic susceptibility, which is given by the reciprocal of the square of elementary charge $e$, i.e.  $\chi_B \rightarrow 1/e^2 $ when $\mu\rightarrow \infty$.

Left--hand panel of Fig. \ref{fig:susceptibility} (a) shows the magnetic susceptibility as functions of temperature, at $eB=0.0~$GeV$^2$ and vanishing $\mu$. The results deduced from PLSM are compared with various lattice simulations (symbols), in which different calculation methods and algorithms are applied. We also compare with the resonance gas model (HRG) model. The recent lattice QCD simulations \cite{Bali:2014kia} (open circle) were estimated by using half--half method in $24^3 \times 32$ lattice (closed triangle) and integral method in $28^3 \times 10$ lattice (open triangle). The diamonds represent lattice simulations for $N_f=2+1$ and when using HISQ/tree action with light quark masses $m_l /m_s=0.05$ and temporal dimension $N_\tau=8$, the lattice results \cite{Levkova:2013qda}. The closed circles stand for results obtained from isotropy lattice \cite{Bali:2013owa}. 

Right--panel of Fig. \ref{fig:susceptibility} (b) depicts the relative permeability with respect to that of the vacuum compared to recent lattice QCD calculations (open triangles) \cite{Bali:2014kia} in a wide range of temperatures, at $eB=0.0~$GeV$^2$ and at vanishing $\mu$. It is apparent that the agreement between PLSM and lattice QCD calculations is excellent.   

Features of PLSM and lattice QCD results can be summarized as follows.
\begin{itemize} 
\item The magnetic susceptibility obtained from the HRG model \cite{Bali:2014kia} (dashed curve) confirms the nature of the QCD matter as dia--magnetized, especially at low temperatures. Here, the free energy density is the sum over all contributions from the colorless hadrons and their resonances contributes to the hadron interactions in order to assure negative magnetic susceptibility \cite{Bali:2014kia}.

\item In PLSM, the free energy density is divided into three terms. The first one is the pure mesonic potential which is obtained from the Lagrangian for the pure gauge. The second one gives the contributions of the quark and antiquarks, which - as the name says - have mesonic fluctuations from the quarks and the antiquarks. The third term represents the interactions of the color charges and the gluons. This means that there two types of contributions to the hadronic fluctuations, while only one type contributes to the gluon interactions.

\item At very low temperatures, the slope of $\chi(T)$ is apparently negative (inside--box in left--hand panel of Fig. \ref{fig:susceptibility}) \cite{Steinert:2013fza}. This gives a signature that the QCD matter is dia--magnetized and also well reproduces the different lattice simulations. When switching to the high temperature regime, i.e. restoring the broken chiral symmetry, we observe a transition between dia-- and para--magnetic properties, where QCD para--magnetism is likely, at high temperature. 

\item We notice that the PLSM results confirm that the strongly interacting QCD matter has para--magnetic properties. The magnetic susceptibility steeply increases when increasing temperatures towards the deconfinement phase--transition. These conclusions have been found in a wide range of temperatures $ 100\leq\,T\,\leq 250$ MeV \cite{Aoki:2009sc,Borsanyi:2010cj}.
\end{itemize}

\subsection{QCD phase--diagram at finite magnetic field \label{phase}}

There are two different mechanisms manifesting the influences of the magnetic field on the QCD phase--diagram. The first one is that the magnetic field improves the QCD phase--transition due to its contributions to the Landau quantizations or the Landau levels. The second one is that the magnetic field contributes to the suppression in the chiral condensate due to the restoration of the broken chiral symmetry. This suppression is known as inverse magnetic catalysis. It is a decrease in the chiral critical temperature with the increase in the magnetic field. 

\begin{figure*}[htb]
\centering
\includegraphics[width=4cm,angle=-90]{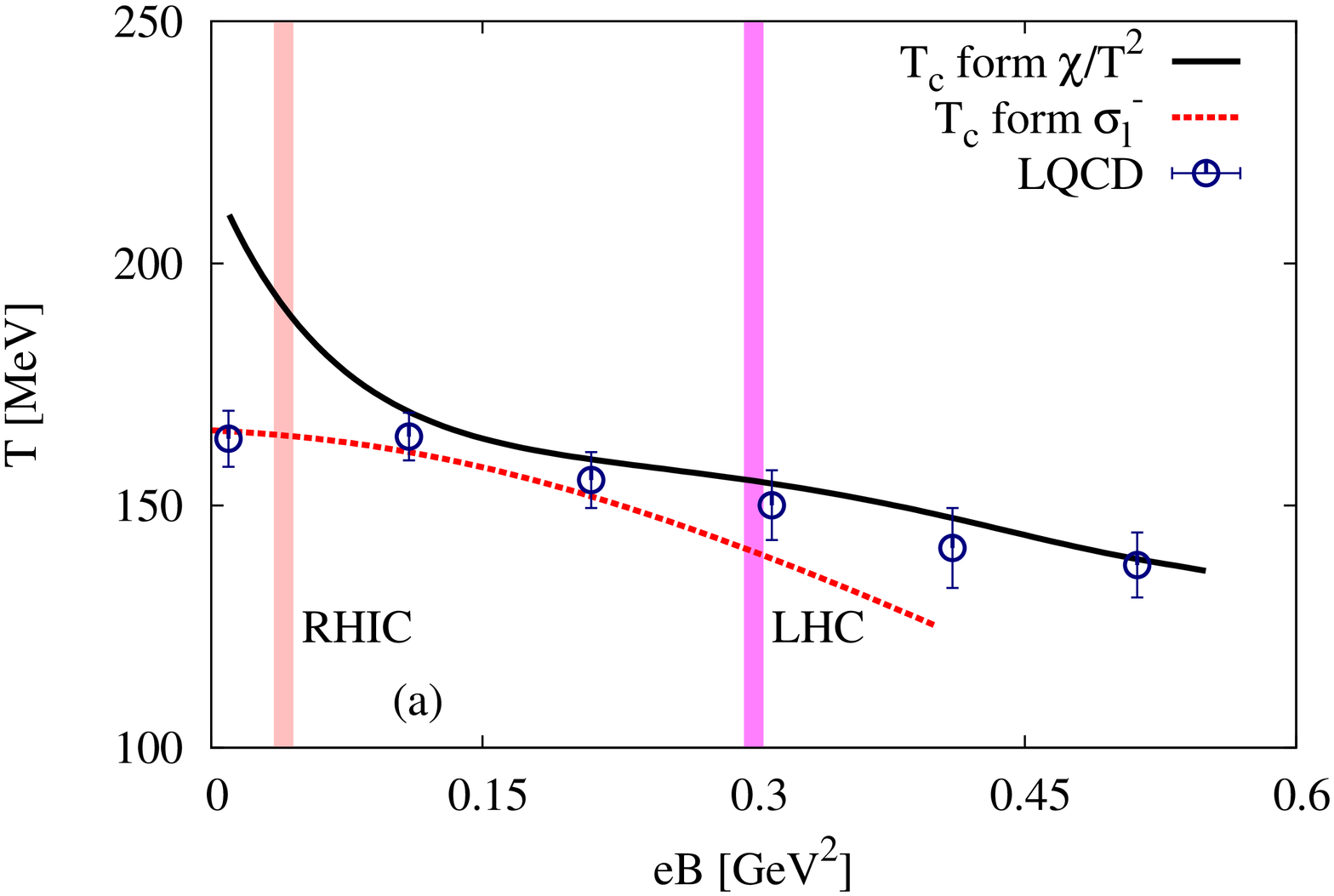}
\includegraphics[width=4cm,angle=-90]{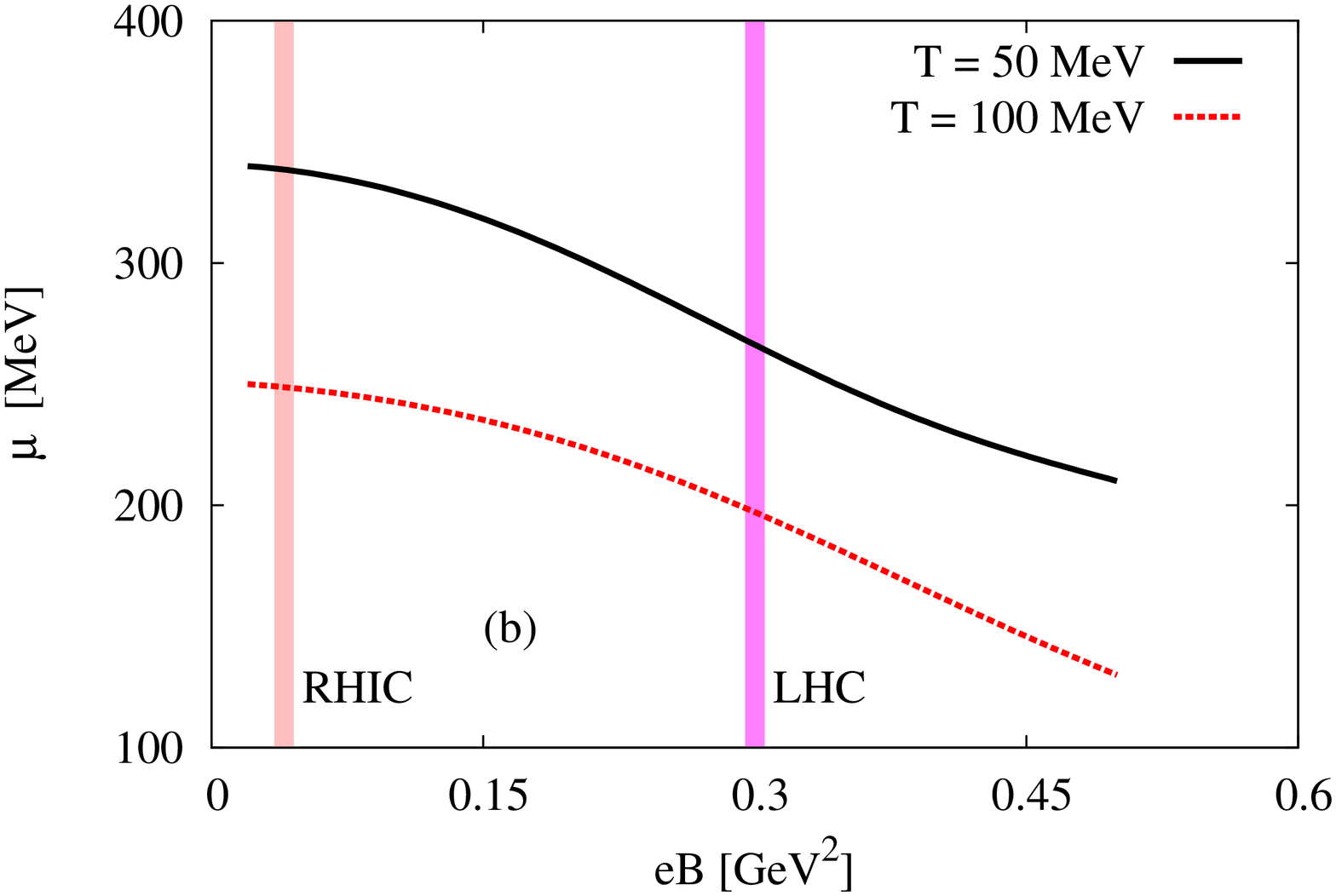}
\includegraphics[width=4cm,angle=-90]{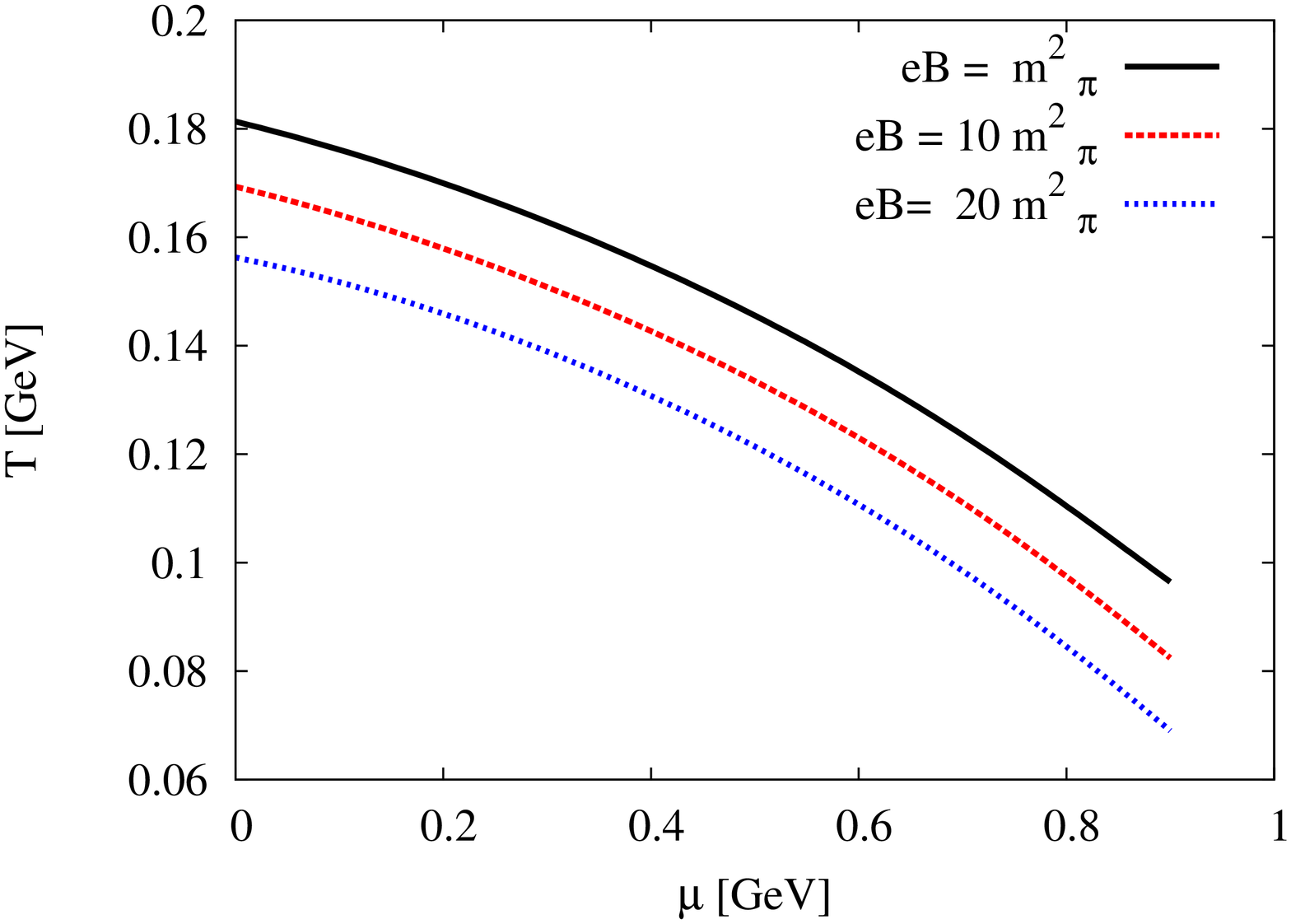}
\caption{Three chiral QCD phase--diagrams. Left--hand panel: critical temperature $T$ to magnetic field $eB$. Middle--panel: critical baryon chemical potential $\mu$ to $eB$. Right--hand panel: $T$ to $\mu$. The vertical bands refer to magnetic fields likely generated, at RHIC and LHC energies. \label{fig:eBdepence} 
}
\end{figure*}

The PLSM has two main types of the order parameters; the chiral condensates, which are connected with two light quarks, $\sigma_l$, and one strange quark, $\sigma_s$, and the Polyakov loop variables, $\phi$ and $\phi^*$. The intersects of $\phi$ and $\phi^*$ with the chiral condensates are used in determining the quasi--critical temperatures (dotted curve) in Fig. \ref{fig:eBdepence}. The solid curve is estimated from higher--order moments of quark multiplicity (not shown in the present paper), where the critical temperatures are estimated at the peak of normalized quark susceptibility $\chi_q/T^2$. From both methods, we find that the critical temperatures decrease with increasing magnetic fields referring to an inverse magnetic catalysis.

In Fig. \ref{fig:eBdepence}, the influences of the magnetic field on the chiral QCD phase--diagram are depicted. Left--hand panel shows the variation of the critical temperature with increasing magnetic field, at almost vanishing $\mu$. The PLSM calculations are compared with recent lattice simulations given (circles with errorbars) \cite{Bali:2014kia}. The vertical bands refer to the magnetic fields generated, at RHIC (orders of per cent of GeV$^2$ or $\sim m_\pi^2$) and LHC energies (orders of per ten GeV$^2$ or $\sim 10-15\, m_\pi^2$). The PLSM calculations take into consideration two methods in order to estimate the critical temperatures. One is based in normalized quark susceptibility $\chi _q/T^2$ and the other one is based on the freeze--out condition $s/T^3$, where $s$ is the entropy density.

There is an excellent agreement, especially at $0\leq\,eB\,$ [GeV$^2$]$\,\leq 0.2$. The solid curve matches well with the lattice simulations, at a wider range of magnetic fields $0.13\leq\,eB\,$ [GeV$^2$]$\,\leq 0.55$. The first method of determining the critical temperature apparently overestimates the lattice results, at low temperature, while the second method slightly underestimates these, especially at high temperatures. We conclude that the chiral magnetic field improves the agreement as it enhances the chiral condensates. This depends also on the type of contributions to the Landau levels introduced to the QCD effective approach. 

The middle--panel draws the dependence of the critical baryon chemical potential on the magnetic fields, at $T=50$ (solid curve) and $100~$MeV (dashed curve). At constant magnetic field as that at RHIC or LHC energy, large critical baryon chemical potential can be reached, even at low temperatures, i.e. $\mu$ decreases with increasing $eB$. The intersection between the deconfinement phase--transition and the light quark chiral condensate is utilized to determine $\mu$, at which the broken chiral symmetry is restored. There is no lattice QCD simulations to compare with. 

The right--hand panel presents $T$-$\mu$ phase--diagram, at $eB=m_{\pi}^2$ (solid curve), $10\, m_{\pi}^2$ (dashed curve) and $20\, m_{\pi}^2$ (dotted curve). In determining the critical temperature, various methods have been utilized, for example, the intersection between the Polyakov loop variables $\phi$ and $\phi^*$, which are related to the deconfinement phase--transition, and the chiral condensates of light and strange quarks, $\sigma_l$ and $\sigma_s$, respectively. The latter is related to the restoration of the broken chiral symmetry. For example, the critical temperature corresponding to the chiral restoration of light quark, $T_c^{\chi _l}$, can be determined by the intersection between $\phi$ and $\sigma_l$, while the critical temperature corresponding to the chiral restoration of strange quark, $T_c^{\chi _s}$, can be defined from the intersection between $\phi^*$ and  $\sigma_s$. We observe that increasing magnetic field enhances the chiral QCD phase--diagram, i.e. the chiral phase--transition takes place at lower temperature. The estimation of freeze--out parameters, $T$ and $\mu$, in dependence on the heavy--ion centralities or the impact parameters allows to analyze the influence of the magnetic field, experimentally \cite{bzdak2012event}. There are various experimental results on chemical and thermal freeze--out as reviewed in ref. \cite{Tawfik:2014eba}.

\begin{figure}[htb]
\centering{
\includegraphics[width=6.cm,angle=0]{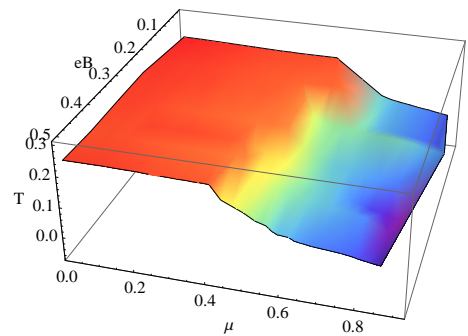}
\caption{As in Fig. \ref{fig:eBdepence} but here the multi--dimensional {\it chemical freeze--out boundary} combines the chemical freeze--out parameters $T$ with the magnetic field $eB$ and the baryon chemical potentila $\mu$. \label{fig:eBdepence2} 
}}
\end{figure}

In Fig. \ref{fig:eBdepence2}, the chemical freeze--out condition $s(T, eB, \mu)/T^3=7$ is implemented in order to estimate the feeze--out temperature $T$. The entropy density is calculated, at different temperatures, baryon chemical potentials, and magnetic fields. At a given value of the entropy density normalized to $T^3$, the related baryon chemical potential, $\mu$, and the corresponding magnetic field, $e B$, are determined. These three parameters are then depicted in Fig. \ref{fig:eBdepence2}. It is a multi--dimensional {\it chemical freeze--out boundary} showing the dependence the freeze--out diagram ($T-\mu$), which can directly be related to the one analyzed from ethe experimental measurements of various particle ratios \cite{Tawfik:2014eba}, on the magnetic field. It is obvious that, at small $\mu$, the effect of the magnetic field is almost negligible. At higher temperatures, the drop in $T_c$ around the chiral phase--transition moves to lower values with increasing $e B$. Again, this refers to an inverse catalysis. At very high temperatures, there a slight increase in $T$ with increasing $eB$. Increasing $e B$ makes the chiral phase--transition smother. It is important to notice that the shape of $T$--$\mu$ phase--diagram looks different from the one at vanishing $eB$ \cite{Tawfik:2014eba}.

\section{Conclusions \label{sec:cncl}}

In relativistic HIC, the off--center motion of spectators in the peripheral collisions and the rapid motion of electric charges generate huge magnetic fields perpendicular to the plane of the motion direction and the electric field. Also, the local imbalance in momenta carried by the colliding nucleons in both peripheral and central collisions generates a huge magnetic field, as well. The response of QCD matter modeled in PLSM to these external magnetic fields such as magnetization, magnetic susceptibility and permeability is determined. The chiral magnetic properties of the QCD phase--diagram are analyzed.
The inclusion of finite magnetic field in PLSM can be accomplished by firstly modifying the dispersion relation of the quarks and antiquarks, so that the dimension of the momentum--space is reduced from three to one and then scaled via quark charge and magnetic field (magnetic catalysis) and secondly Landau quantization should be integrated in.
The latter modifies the dispersion relation by a quantization number; the Landau quantum number. $\sigma$, which is related to the spin quantum number and to the masses of quark flavors. Both chiral condensates and deconfinement order--parameters have been analyzed in a wide range of temperatures, baryon chemical potentials, magnetic fields, so that the chiral QCD phase--diagram could be mapped out in various directions.  

The magnetization remarkably affects the thermodynamic properties of the QCD matter. The PLSM results give an evident on paramagnetic features of the hot QCD matter. The magnetic susceptibility, at low temperatures, is negative indicating that the QCD matter is dia--magnetized. At higher temperatures regime, i.e. restoring the broken chiral symmetry, there is a transition from dia-- to para--magnetic properties. The permeability is a characterizing property of a magnetic material measuring the ability to create magnetic field and to store magnetic potential energy. The latter is proportionally constant for magnetic flux, which is produced from the influences of the magnetic field. At low temperatures (hadron phase) the QCD permeability is small but rapidly increases around the deconfinement phase--transition. At high temperatures (parton phase) the QCD permeability is large. We conclude that in a wide range of temperatures, the magnetic permeability normalized to the vacuum value agrees well with the lattice QCD simulations, so that this quantity can be utilized as a magnetic order--parameter. 

To estimate the variation of the chiral critical temperature with the magnetic field, the PLSM results are confronted to the recent lattice QCD data. We found that the chiral critical temperature decreases as the magnetic field increases. We conclude that the magnetic catalysis of the thermal QCD medium is inverse and an inverse interrelation between the chiral chemical potential and the magnetic field is obtained. The chiral phase--diagram is shifted to lower temperatures due to the increase in the magnetic field. This result is confirmed by two different methods, one from the thermal and dense phase--transition of the PLSM parameters and the another one by applying a condition from freeze--out parameters $s/T^3$.

%
%
\bibliographystyle{aip}
\bibliography{Refs} 

\def\germ{\frak} \def\scr{\cal} \ifx\documentclass\undefinedcs
  \def\bf{\fam\bffam\tenbf}\def\rm{\fam0\tenrm}\fi 
  \def\defaultdefine#1#2{\expandafter\ifx\csname#1\endcsname\relax
  \expandafter\def\csname#1\endcsname{#2}\fi} \defaultdefine{Bbb}{\bf}
  \defaultdefine{frak}{\bf} \defaultdefine{=}{\B} 
  \defaultdefine{mathfrak}{\frak} \defaultdefine{mathbb}{\bf}
  \defaultdefine{mathcal}{\cal} \defaultdefine{implies}{\Rightarrow}
  \defaultdefine{beth}{BETH}\defaultdefine{cal}{\bf} \def\bbfI{{\Bbb I}}
  \def\mbox{\hbox} \def\text{\hbox} \def\om{\omega} \def\Cal#1{{\bf #1}}
  \def\pcf{pcf} \defaultdefine{cf}{cf} \defaultdefine{reals}{{\Bbb R}}
  \defaultdefine{real}{{\Bbb R}} \def\restriction{{|}} \def\club{CLUB}
  \def\w{\omega} \def\exist{\exists} \def\se{{\germ se}} \def\bb{{\bf b}}
  \def\equivalence{\equiv} \let\lt< \let\gt>
\begin{thebibliography}{10}

\bibitem{Hagedorn:1965st}
R.~Hagedorn,
\newblock Nuovo Cim. Suppl. {\bf 3}, 147 (1965).

\bibitem{Hagedorn:1984hz}
R.~Hagedorn,
\newblock Lect. Notes Phys. {\bf 221}, 53 (1985).

\bibitem{Cabibbo:1975ig}
N.~Cabibbo and G.~Parisi,
\newblock Phys. Lett. {\bf 59B}, 67 (1975).

\bibitem{Collins:1974ky}
J.~C. Collins and M.~J. Perry,
\newblock Phys. Rev. Lett. {\bf 34}, 1353 (1975).

\bibitem{Baym:2001in}
G.~Baym,
\newblock Nucl. Phys. {\bf A698}, XXIII (2002).

\bibitem{fukushima2008chiral}
K.~Fukushima,
\newblock Journal of Physics G: Nuclear and Particle Physics {\bf 35}, 104020
  (2008).

\bibitem{huang2016electromagnetic}
X.-G. Huang,
\newblock Reports on Progress in Physics {\bf 79}, 076302 (2016).

\bibitem{Vachaspati:1991nm}
T.~Vachaspati,
\newblock Phys. Lett. {\bf B265}, 258 (1991).

\bibitem{bzdak2012event}
A.~Bzdak and V.~Skokov,
\newblock Physics Letters B {\bf 710}, 171 (2012).

\bibitem{McLerran:2013hla}
L.~McLerran and V.~Skokov,
\newblock Nucl. Phys. {\bf A929}, 184 (2014).

\bibitem{Tuchin:2013apa}
K.~Tuchin,
\newblock Phys. Rev. {\bf C88}, 024911 (2013).

\bibitem{Tuchin:2014hza}
K.~Tuchin,
\newblock Int. J. Mod. Phys. {\bf E23}, 1430001 (2014).

\bibitem{Bali:2011qj}
G.~S. Bali et~al.,
\newblock JHEP {\bf 02}, 044 (2012).

\bibitem{Levkova:2013qda}
L.~Levkova and C.~DeTar,
\newblock Phys. Rev. Lett. {\bf 112}, 012002 (2014).

\bibitem{Bonati:2013vba}
C.~Bonati, M.~D'Elia, M.~Mariti, F.~Negro, and F.~Sanfilippo,
\newblock Phys. Rev. {\bf D89}, 054506 (2014).

\bibitem{Bruckmann:2013oba}
F.~Bruckmann, G.~Endrodi, and T.~G. Kovacs,
\newblock JHEP {\bf 04}, 112 (2013).

\bibitem{Bali:2013txa}
G.~S. Bali, F.~Bruckmann, G.~Endrödi, and A.~Schäfer,
\newblock PoS {\bf LATTICE2013}, 182 (2014).

\bibitem{Endrodi:2013cs}
G.~Endrödi,
\newblock JHEP {\bf 04}, 023 (2013).

\bibitem{Bhattacharyya:2015pra}
A.~Bhattacharyya, S.~K. Ghosh, R.~Ray, and S.~Samanta,
\newblock EPL {\bf 115}, 62003 (2016).

\bibitem{Ratti:2005jh}
C.~Ratti, M.~A. Thaler, and W.~Weise,
\newblock Phys. Rev. {\bf D73}, 014019 (2006).

\bibitem{Ferreira:2015gxa}
M.~Ferreira, P.~Costa, C.~Providência, O.~Lourenço, and T.~Frederico,
\newblock {Inverse Magnetic Catalysis in hot quark matter within (P)NJL
  models},
\newblock in {\em {Proceedings, Compact Stars in the QCD Phase Diagram IV
  (CSQCD IV): Prerow, Germany, September 26-30, 2014}}, 2015.

\bibitem{Ferreira:2014kpa}
M.~Ferreira, P.~Costa, O.~Lourenço, T.~Frederico, and C.~Providência,
\newblock Phys. Rev. {\bf D89}, 116011 (2014).

\bibitem{Farias:2014eca}
R.~L.~S. Farias, K.~P. Gomes, G.~I. Krein, and M.~B. Pinto,
\newblock Phys. Rev. {\bf C90}, 025203 (2014).

\bibitem{Chao:2013qpa}
J.~Chao, P.~Chu, and M.~Huang,
\newblock Phys. Rev. {\bf D88}, 054009 (2013).

\bibitem{Mei:2020jzn}
J.~Mei and S.~Mao,
\newblock Phys. Rev. {\bf D102}, 114035 (2020).

\bibitem{Mao:2016fha}
S.~Mao,
\newblock Phys. Lett. {\bf B758}, 195 (2016).

\bibitem{Andersen:2021lnk}
J.~O. Andersen,
\newblock (arXiv:2102.13165 [hep-ph]).

\bibitem{Shovkovy:2012zn}
I.~A. Shovkovy,
\newblock Lect. Notes Phys. {\bf 871}, 13 (2013).

\bibitem{Andersen:2014xxa}
J.~O. Andersen, W.~R. Naylor, and A.~Tranberg,
\newblock Rev. Mod. Phys. {\bf 88}, 025001 (2016).

\bibitem{Kharzeev:2012ph}
D.~E. Kharzeev, K.~Landsteiner, A.~Schmitt, and H.-U. Yee,
\newblock Lect. Notes Phys. {\bf 871}, 1 (2013).

\bibitem{Gatto:2012sp}
R.~Gatto and M.~Ruggieri,
\newblock Lect. Notes Phys. {\bf 871}, 87 (2013).

\bibitem{gell1960axial}
M.~Gell-Mann and M.~L{\'e}vy,
\newblock Il Nuovo Cimento (1955-1965) {\bf 16}, 705 (1960).

\bibitem{Lenaghan:1999si}
J.~T. Lenaghan and D.~H. Rischke,
\newblock J. Phys. {\bf G26}, 431 (2000).

\bibitem{Petropoulos:1998gt}
N.~Petropoulos,
\newblock J. Phys. {\bf G25}, 2225 (1999).

\bibitem{levy1967currents}
M.~L{\'e}vy,
\newblock Il Nuovo Cimento A (1965-1970) {\bf 52}, 23 (1967).

\bibitem{hu1974chiral}
B.~Hu,
\newblock Physical Review D {\bf 9}, 1825 (1974).

\bibitem{geddes1980spin}
H.~Geddes,
\newblock Physical Review D {\bf 21}, 278 (1980).

\bibitem{Tawfik:2014uka}
A.~Tawfik, N.~Magdy, and A.~Diab,
\newblock Phys. Rev. {\bf C89}, 055210 (2014).

\bibitem{Tawfik:2014gga}
A.~N. Tawfik and A.~M. Diab,
\newblock Phys. Rev. {\bf C91}, 015204 (2015).

\bibitem{Tawfik:2019rdd}
A.~N. Tawfik, A.~M. Diab, and M.~T. Hussein,
\newblock Chin. Phys. {\bf C43}, 034103 (2019).

\bibitem{Tawfik:2016lih}
A.~N. Tawfik, A.~M. Diab, N.~Ezzelarab, and A.~G. Shalaby,
\newblock Adv. High Energy Phys. {\bf 2016}, 1381479 (2016).

\bibitem{Tawfik:2016gye}
A.~N. Tawfik, A.~M. Diab, and M.~T. Hussein,
\newblock J. Phys. {\bf G45}, 055008 (2018).

\bibitem{AbdelAalDiab:2016rje}
A.~M. Abdel Aal~Diab, A.~N. Tawfik, and M.~T. Hussein,
\newblock {Electromagnetic Effects on Strongly Interacting QCD-Matter},
\newblock 2016.

\bibitem{Tawfik:2017cdx}
A.~N. Tawfik, A.~M. Diab, and M.~T. Hussein,
\newblock J. Exp. Theor. Phys. {\bf 126}, 620 (2018).

\bibitem{Tawfik:2016edq}
A.~N. Tawfik, A.~M. Diab, and M.~T. Hussein,
\newblock Int. J. Mod. Phys. {\bf A31}, 1650175 (2016).

\bibitem{Tawfik:2016ihn}
A.~N. Tawfik, A.~M. Diab, and T.~M. Hussein,
\newblock Int. J. Adv. Res. Phys. Sci. {\bf 3}, 4 (2016).

\bibitem{Diab:2016iig}
A.~M. Diab, A.~I. Ahmadov, A.~N. Tawfik, and E.~A.~E. Dahab,
\newblock PoS {\bf ICHEP2016}, 634 (2016).

\bibitem{AbdelAalDiab:2018hrx}
A.~M. Abdel Aal~Diab and A.~N. Tawfik,
\newblock EPJ Web Conf. {\bf 177}, 09005 (2018).

\bibitem{Scavenius:2000qd}
O.~Scavenius, A.~Mocsy, I.~N. Mishustin, and D.~H. Rischke,
\newblock Phys. Rev. {\bf C64}, 045202 (2001).

\bibitem{Koch:1997ei}
V.~Koch,
\newblock Int. J. Mod. Phys. {\bf E6}, 203 (1997).

\bibitem{gasiorowicz1969effective}
S.~Gasiorowicz and D.~A. Geffen,
\newblock Reviews of Modern Physics {\bf 41}, 531 (1969).

\bibitem{Ko:1994en}
P.~Ko and S.~Rudaz,
\newblock Phys. Rev. {\bf D50}, 6877 (1994).

\bibitem{Parganlija:2008jf}
D.~Parganlija, F.~Giacosa, and D.~H. Rischke,
\newblock PoS {\bf CONFINEMENT8}, 070 (2008).

\bibitem{Pisarski:1994yp}
R.~D. Pisarski,
\newblock {Applications of chiral symmetry},
\newblock in {\em {Workshop on Finite Temperature QCD and Quark - Gluon
  Transport Theory Wuhan, China, April 18-26, 1994}}, 1994.

\bibitem{Kovacs:2013xca}
P.~Kovács and G.~Wolf,
\newblock Acta Phys. Polon. Supp. {\bf 6}, 853 (2013).

\bibitem{Parganlija:2012fy}
D.~Parganlija, P.~Kovacs, G.~Wolf, F.~Giacosa, and D.~H. Rischke,
\newblock Phys. Rev. {\bf D87}, 014011 (2013).

\bibitem{Fariborz:2008bd}
A.~H. Fariborz, R.~Jora, and J.~Schechter,
\newblock Phys. Rev. {\bf D77}, 094004 (2008).

\bibitem{Rosenzweig:1979ay}
C.~Rosenzweig, J.~Schechter, and C.~G. Trahern,
\newblock Phys. Rev. {\bf D21}, 3388 (1980).

\bibitem{weinberg1975u}
S.~Weinberg,
\newblock Physical Review D {\bf 11}, 3583 (1975).

\bibitem{Schaefer:2008hk}
B.-J. Schaefer and M.~Wagner,
\newblock Phys. Rev. {\bf D79}, 014018 (2009).

\bibitem{Lenaghan:2000ey}
J.~T. Lenaghan, D.~H. Rischke, and J.~Schaffner-Bielich,
\newblock Phys. Rev. {\bf D62}, 085008 (2000).

\bibitem{Mao:2009aq}
H.~Mao, J.~Jin, and M.~Huang,
\newblock J. Phys. {\bf G37}, 035001 (2010).

\bibitem{Polyakov:1978vu}
A.~M. Polyakov,
\newblock Phys. Lett. {\bf 72B}, 477 (1978).

\bibitem{Roessner:2006xn}
S.~Roessner, C.~Ratti, and W.~Weise,
\newblock Phys. Rev. {\bf D75}, 034007 (2007).

\bibitem{Fukushima:2008wg}
K.~Fukushima,
\newblock Phys. Rev. {\bf D77}, 114028 (2008),
\newblock [Erratum: Phys. Rev.D78,039902(2008)].

\bibitem{Schaefer:2007pw}
B.-J. Schaefer, J.~M. Pawlowski, and J.~Wambach,
\newblock Phys. Rev. {\bf D76}, 074023 (2007).

\bibitem{langelage2011centre}
J.~Langelage, S.~Lottini, and O.~Philipsen,
\newblock Journal of High Energy Physics {\bf 2011}, 57 (2011).

\bibitem{Lo:2013hla}
P.~M. Lo, B.~Friman, O.~Kaczmarek, K.~Redlich, and C.~Sasaki,
\newblock Phys. Rev. {\bf D88}, 074502 (2013).

\bibitem{Menezes:2009uc}
D.~P. Menezes, M.~Benghi~Pinto, S.~S. Avancini, and C.~Providencia,
\newblock Phys. Rev. {\bf C80}, 065805 (2009).

\bibitem{Wen:2016atg}
X.-J. Wen and J.-J. Liang,
\newblock Phys. Rev. {\bf D94}, 014005 (2016).

\bibitem{Avancini:2011zz}
S.~S. Avancini, D.~P. Menezes, and C.~Providencia,
\newblock Phys. Rev. {\bf C83}, 065805 (2011).

\bibitem{Boomsma:2009yk}
J.~K. Boomsma and D.~Boer,
\newblock Phys. Rev. {\bf D81}, 074005 (2010).

\bibitem{Tawfik:2015tga}
A.~N. Tawfik and N.~Magdy,
\newblock Phys. Rev. {\bf C91}, 015206 (2015).

\bibitem{Kovacs:2006ym}
P.~Kovacs and Z.~Szep,
\newblock Phys. Rev. {\bf D75}, 025015 (2007).

\bibitem{Tawfik:2019tkp}
A.~N. Tawfik, A.~M. Diab, M.~T. Ghoneim, and H.~Anwer,
\newblock Int. J. Mod. Phys. {\bf A34}, 1950199 (2019).

\bibitem{Fukushima:2008xe}
K.~Fukushima, D.~E. Kharzeev, and H.~J. Warringa,
\newblock Phys. Rev. {\bf D78}, 074033 (2008).

\bibitem{Fukushima:2009ft}
K.~Fukushima, D.~E. Kharzeev, and H.~J. Warringa,
\newblock Nucl. Phys. {\bf A836}, 311 (2010).

\bibitem{Dumitru:2005ng}
A.~Dumitru, R.~D. Pisarski, and D.~Zschiesche,
\newblock Phys. Rev. {\bf D72}, 065008 (2005).

\bibitem{Aad:2012ew}
G.~Aad et~al.,
\newblock Phys. Rev. Lett. {\bf 110}, 022301 (2013).

\bibitem{Tawfik:2012ty}
A.~Nasser~Tawfik and H.~Magdy,
\newblock Int. J. Mod. Phys. {\bf A29}, 1450152 (2014).

\bibitem{Bali:2014kia}
G.~S. Bali, F.~Bruckmann, G.~Endrödi, S.~D. Katz, and A.~Schäfer,
\newblock JHEP {\bf 08}, 177 (2014).

\bibitem{Kamikado:2014bua}
K.~Kamikado and T.~Kanazawa,
\newblock JHEP {\bf 01}, 129 (2015).

\bibitem{Tawfik:2016cot}
A.~N. Tawfik,
\newblock Indian J. Phys. {\bf 91}, 93 (2017).

\bibitem{Bali:2013owa}
G.~S. Bali, F.~Bruckmann, G.~Endrodi, and A.~Schafer,
\newblock Phys. Rev. Lett. {\bf 112}, 042301 (2014).

\bibitem{Steinert:2013fza}
T.~Steinert and W.~Cassing,
\newblock Phys. Rev. {\bf C89}, 035203 (2014).

\bibitem{Aoki:2009sc}
Y.~Aoki et~al.,
\newblock JHEP {\bf 06}, 088 (2009).

\bibitem{Borsanyi:2010cj}
S.~Borsanyi et~al.,
\newblock JHEP {\bf 11}, 077 (2010).

\bibitem{Tawfik:2014eba}
A.~N. Tawfik,
\newblock Int. J. Mod. Phys. {\bf A29}, 1430021 (2014).

\end{thebibliography}
\end{document}